\documentclass[aps,pra,twocolumn,showpacs,superscriptaddress]{revtex4}

\usepackage{graphicx,color}
\usepackage{dcolumn}
\usepackage{bm}

\begin{document}


\title{Coherent bremsstrahlung, coherent pair production, birefringence and 
polarimetry in the 20-170 GeV energy range using aligned crystals}



\author{A.~Apyan}
\altaffiliation[Now at: ]{Northwestern University, Evanston, IL, USA}
\affiliation{Yerevan Physics Institute, Yerevan, Armenia}

\author{R.O.~Avakian}
\affiliation{Yerevan Physics Institute, Yerevan, Armenia}

\author{B.~Badelek}
\affiliation{Uppsala University, Uppsala, Sweden}

\author{S.~Ballestrero}
\affiliation{INFN and University of Firenze, Firenze, Italy and CERN, Geneva, Switzerland}
\altaffiliation[Now at: ]{University of the Witwatersrand,
Johannesburg, South Africa}

\author{C.~Biino}
\affiliation{INFN and University of Torino, Torino, Italy}
\affiliation{CERN, Geneva, Switzerland}

\author{I.~Birol}
\affiliation{Northwestern University, Evanston, IL, USA}

\author{P.~Cenci}
\affiliation{INFN, Perugia, Italy}

\author{S.H.~Connell}
\affiliation{School of Physics - University of the Witwatersrand,
Johannesburg, South Africa}

\author{S.~Eichblatt}
\affiliation{Northwestern University, Evanston, IL, USA}

\author{T.~Fonseca}
\affiliation{Northwestern University, Evanston, IL, USA}

\author{A.~Freund}
\affiliation{ESRF, Grenoble, France}

\author{B.~Gorini}
\affiliation{CERN, Geneva, Switzerland}

\author{R.~Groess}
\affiliation{Schonland Research Centre - University of the Witwatersrand,
Johannesburg, South Africa}

\author{K.~Ispirian}
\affiliation{Yerevan Physics Institute, Yerevan, Armenia}

\author{T.J.~Ketel}
\affiliation{NIKHEF, Amsterdam, The Netherlands}

\author{Yu.V.~Kononets}
\affiliation{Kurchatov Institute, Moscow, Russia}

\author{A.~Lopez}
\affiliation{University of Santiago de Compostela, Santiago de Compostela,
Spain}

\author{A.~Mangiarotti}
\affiliation{INFN and University of Firenze, Firenze, Italy}

\author{B.~van~Rens}
\affiliation{NIKHEF, Amsterdam, The Netherlands}

\author{J.P.F.~Sellschop}
\altaffiliation[Deceased]{}
\affiliation{Schonland Research Centre - University of the Witwatersrand,
Johannesburg, South Africa}

\author{M.~Shieh}
\affiliation{Northwestern University, Evanston, IL, USA}

\author{P.~Sona}
\affiliation{INFN and University of Firenze, Firenze, Italy}

\author{V.~Strakhovenko}
\affiliation{Institute of Nuclear Physics, Novosibirsk, Russia}

\author{E.~Uggerh{\o}j}
\thanks{Co-Spokeperson}
\affiliation{Institute for Storage Ring Facilities, University of Aarhus,
Denmark}

\author{U.I.~Uggerh{\o}j}
\affiliation{University of Aarhus, Aarhus, Denmark}

\author{G.~Unel}
\affiliation{University of California at Irvine, Phys. Dept., and CERN, Phys. Dept.}

\author{M.~Velasco}
\thanks{Co-Spokeperson}
\altaffiliation[Now at: ]{Northwestern University, Evanston, IL, USA}
\affiliation{CERN, Geneva, Switzerland}

\author{Z.Z.~Vilakazi}
\altaffiliation[Now at: ]{University of Cape Town, Cape Town, South Africa}

\affiliation{Schonland Research Centre - University of the Witwatersrand,
Johannesburg, South Africa}

\author{O.~Wessely}
\affiliation{Uppsala University, Uppsala, Sweden}

\collaboration{The NA59 Collaboration}

\noaffiliation

\date{\today}

\begin{abstract}
The processes of coherent bremsstrahlung (CB) and coherent pair production 
(CPP) based on aligned crystal targets have been studied in the energy range 
20-170 GeV. The experimental arrangement allowed for measurements
of single photon properties of these phenomena including
their polarization dependences. This is significant as 
the theoretical description of CB and CPP is an area of active theoretical 
debate and development. With the theoretical approach used in this paper both
the measured cross sections and polarization observables
are predicted very well. This indicates a proper understanding
of CB and CPP up to energies of 170 GeV.
Birefringence in CPP on aligned crystals is applied to
determine the polarization parameters in our measurements.
New technologies for high energy photon beam optics
including phase plates and polarimeters for linear and 
circular polarization are demonstrated in this experiment.
Coherent bremsstrahlung for the strings-on-strings (SOS)
orientation yields a larger enhancement for hard photons
than CB for the channeling orientations of the crystal. 
Our measurements and our calculations indicate low photon
polarizations for the high energy SOS photons.
\end{abstract}

\pacs{03.65.Sq, 61.85.+p, 78.20.Fm, 78.70.-g, 95.75.Hi}

\maketitle

\section{\label{sec:intro}Introduction}

The demand for high energy circularly polarized photon beams has increased 
with the need to study gluon related features of the nucleon. The so-called 
"spin crisis of the nucleon" and its connection to the gluon polarization has 
attracted much attention~\cite{compass}. For example, experiments to determine 
the gluon spin density of the nucleon~\cite{compass,ric,bosted} from polarized 
virtual photon-gluon fusion, and polarized virtual photoproduction of high 
transverse momentum mesons~\cite{afanas}. Future experiments will require 
intense high energy photon beams with a high degree of circular polarization. 
A well known method to produce circularly polarized photons is the 
interaction of longitudinally polarized electrons with crystalline media, 
where the emitted photons are circularly polarized due to conservation of 
angular momentum~\cite{olsen}. Specifically, theoretical 
calculations~\cite{nadz,armen} predict that coherent bremsstrahlung 
(CB) and channeling radiation (CR) in crystals by longitudinally polarized 
electrons are also circularly polarized, and this can be used to enhance the 
number of high energy circularly polarized photons. The subject of interactions of relativistic particles with strong crystalline fields has 
been recently reviewed~\cite{ulrik}. Currently, the highest 
energy polarized electron beam available is 45 GeV~\cite{slac1,slac2}. 
Photons that will allow $\gamma g\rightarrow c\bar{c}$ in $\Lambda_c D$ 
production with a four-momentum fraction of the gluon of $\eta$ have a 
threshold energy of 9.2/$\eta$~GeV. Therefore, the available polarized 
electron beams cannot produce polarized photons that are sufficiently 
energetic, that is above 92~GeV, to investigate the gluon spin contribution 
to the proton by the above mentioned reaction for $\eta$ values of about 0.1.

Unpolarized electron beams are available to much higher energies, for example, 
energies of up to 250\,GeV at CERN and 125\,GeV at FNAL. Linearly polarized 
photons may be produced from such beams by CB. It is therefore of interest to 
investigate the possible conversion of this linear polarization to circular, 
and to develop polarimetry techniques at these very high photon energies. 
CB radiation differs from incoherent bremsstrahlung (ICB) in an amorphous 
target in that the cross section is substantially enhanced with relatively 
sharp peaks in the photon spectrum. The position of these peaks can be tuned 
by adjusting the electron beam incidence angle with respect to the major 
planes of the lattice. New features of coherent high energy photon emission 
develop at higher electron energies. For certain geometries for the incident 
electron beam with respect to the aligned crystal target, the so-called "strong 
field" effects become important. A special case is found if 
the electron beam is incident very close to the 
plane (within the planar channeling critical angle) and also closely aligned 
to a major axis (but beyond the axial channeling critical angle). Here the 
electrons interact dominantly with successive atomic strings in the plane. 
This orientation was aptly described by the term "string-of-strings" (SOS) by 
Lindhard, a pioneer of beam-crystal phenomena~\cite{lindhard}. The 
polarization features of this SOS radiation require further investigation. 
In conclusion, a study of the above mentioned phenomena constitutes an 
opportunity to benchmark the latest theoretical approaches that describe CB 
and also the related process of CPP at these energies.

Accordingly, this paper has three distinct sections. The first section studies 
birefringent effects in CPP by photons in the 20-170 GeV energy range incident 
on aligned crystals. As a by-product, a new crystal polarimetry technique is 
established. The second section extends the investigation of birefringence by 
aligned crystals and demonstrates the conversion of linear polarization to 
circular polarization for the CB photons. The crystal polarimetry technique 
is here extended to quantify also circular polarization. The third section 
addresses the issue of the polarization of SOS radiation. Also in this 
section it is demonstrated that the theory which is discussed accurately 
describes the observations. The theoretical calculations cover the cross 
sections for CB and SOS radiation for the photon generation, the cross section 
for CPP for the polarimetry, and the linear to circular polarization 
conversion. Simulations based on the theoretical calculations therefore 
predict the measured polarization observables. The good agreement between our 
measurements and the simulations indicate that, even for the strong field 
case, the theoretical description is reliable.

This work focuses both on cross sections and polarization phenomena in CB and 
CPP at high energies in oriented single crystals. The CB and CPP theories are 
constructed in the framework of the first Born approximation in the crystal 
potential. These theories are well established and were experimentally 
investigated for up to a few tens of GeV electrons and photons. The 
theoretical description of those phenomena in oriented single crystals becomes 
more complicated at higher energies. The processes have strong angular and 
energy dependence and the validity conditions of the Born approximation no 
longer hold at very high energies and small incidence angles with respect to 
the crystal axes and planes. The onset of this problem for the description of 
radiation emission and pair production (PP) has the characteristic angle 
$\theta_v=U_0/m$~\cite{baier} where $U_0$ is the plane potential well depth, 
$m$ is the electron rest mass and $\hbar=c=1$. The radiation and pair 
production processes can be described by the CB and CPP theory for the 
incidence angles with respect to the crystal axes/planes 
$\theta \gg \theta_v$. For angles $\theta \sim \theta_v$ and 
$\theta < \theta_v$ a different approach, known as the quasi classical 
description is used. In this approach the general theory of radiation and 
pair production is developed based on the quasi classical operator 
method~\cite{baier}.

\section{\label{sec:production}Production of high energy photon beams}

As described below, the so-called point effect (PE) orientation of the crystal 
was used in the first section of the experiment, where linear polarization 
studies of high energy CB photons were performed and birefringent effects 
in pair production on aligned crystals were studied. This section of the 
experiment also leads to a new polarimetry technique~\cite{na59-1}.
                                                                                
The same orientation was also used in the second section of the experiment, 
where the conversion of the linear polarization to circular polarization 
induced by a birefringent effect in an aligned single crystal was 
studied~\cite{na59-2}.
                                                                                
For the third section of the experiment, the crystal orientation appropriate 
for SOS radiation was used~\cite{na59-3}. This radiation production scenario 
is also treated below.

\subsection{\label{subsec:CB-prod}Linearly polarized CB photons}

In the production of photon beams, single crystals can play an important role 
by exploiting coherent and strong field effects that arise for oriented 
incidence in the interaction of radiation and matter in crystalline 
materials~\cite{kirsebom01}. The CB method is a well established technique 
for obtaining linearly polarized photons starting from unpolarized 
electrons~\cite{feretti, termisha, uberall, diambrini}. An electron impinging 
on a crystal will interact coherently with the electric fields of the atoms 
in aligned crystal planes. If the Laue condition is satisfied, the 
bremsstrahlung photons will be emitted at specific energies corresponding 
to the selected vectors of the reciprocal lattice. In the so-called PE 
orientation of the crystal the direction of the electron beam has a small 
angle with respect to a chosen crystallographic plane and a relatively large 
angle with the crystallographic axes that are in that plane. For this PE 
orientation of the single crystal only one reciprocal lattice vector 
contributes to the CB cross section. The CB radiation  from a crystal aligned 
in this configuration is more intense than the ICB radiation in amorphous 
media and a high degree of linear polarization can be 
achieved~\cite{termisha}. The maximum polarization and the maximum peak 
intensity occur at the same photon energy, and this energy can be selected by 
choosing the orientation of the lattice planes with respect to the incoming 
electron beam. This property has been used previously to achieve photon beams 
with up to 70\% linear polarization starting from  6\,GeV 
electrons~\cite{6gev}, and up to 60\% linear polarization starting from  
80~GeV electrons~\cite{omega}.

The emission mechanism of the high energy photons (CB) is connected to the 
periodic structure of the crystal~\cite{termisha}. The peak energy of the CB 
photons, $E_\gamma$, is determined from the condition (the system of units 
used here has $\hbar={\rm c}=1$ ),
\begin{equation}
\frac{1}{|q_{\Vert}|} = 2 \lambda_c \gamma \frac{E_0-E_\gamma}{E_\gamma}~,
\label{eq:E-peak}
\end{equation}
where $|q_{\Vert}|$ is the component of the recoil momentum of the nucleus 
parallel to the initial electron velocity and the other symbols have their 
usual meanings. Recall, in a crystal possible values of $\mathbf{q}$, from 
which the contribution to the coherent radiation comes, are discrete: 
$\mathbf{q}=\mathbf{g}$~\cite{termisha}, where $\mathbf{g}$ is a reciprocal 
lattice vector of the crystal. The minimal reciprocal lattice vector giving 
rise to the main CB peak is given by
\begin{equation}
|g_{\Vert}|_{min} = \frac{2\pi}{d}\Theta.
\end{equation}

For the PE orientation, $d$ is the interplanar distance and $\Theta=\psi$, 
the electron incident angle with respect to the plane.

The position of the hard photon peak can be selected by simultaneous solution 
of the last two equations
\begin{equation}
\Theta =\frac{d}{4\pi\gamma\lambda_c}\frac{E_{\gamma}}{E_0-E_{\gamma}}.
\end{equation}

The coherence length determines the effective longitudinal dimension of the 
interaction region for the phase coherence of the radiation process:
\begin{equation}
l_{coh} = \frac{1}{|q_{\Vert}|}.
\end{equation}

\begin{figure}[htbp]
\includegraphics[width=0.48\textwidth]{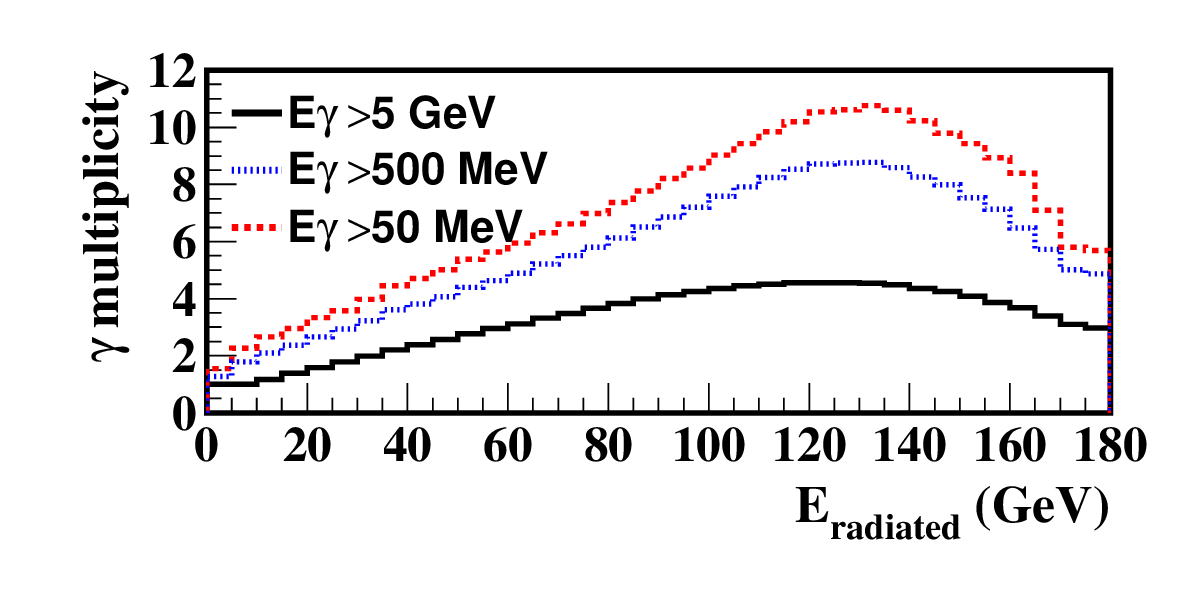}
\caption{\label{fig:multip} Monte Carlo prediction for photon multiplicity 
vs total radiated  energy  using different photon energy cut-off values.}
\end{figure}

The relative merits of different single crystals as CB radiators have been 
investigated in the past~\cite{xtalprops}. The silicon crystal stands out as 
a good choice due to its availability, ease of growth, and low mosaic spread (high lattice quality). 
A Si crystal thickness of 1.5~cm was selected to achieve a relatively low 
photon multiplicity and reasonable photon emission rate. This multiplicity is 
shown in Fig.~\ref{fig:multip} to reside dominantly in lower energy photons.

The multiplicity is the weighted average of the number of photons per incident 
electron:


\begin{equation}
M = \frac{\sum \limits_{i=1}^{n} i N_i}{\sum \limits_{i=1}^{n} N_i}
\end{equation}
where $i = 1, 2, 3, \ldots n$ is the number of radiated photons 
(with an energy above the $E_{\gamma}$ threshold) per electron and 
$N_i$ is the number of electrons radiating the number of 
photons, $i$. For example, $N_1$ is the number of electrons radiating only one 
photon, $N_2$ is the number of electrons radiating two photons, etc. 
$ N = \sum \limits_{i} N_i$ is the total number of primary electrons.

\vspace{1cm}

\begin{figure}[htbp]
\includegraphics[width=0.45\textwidth]{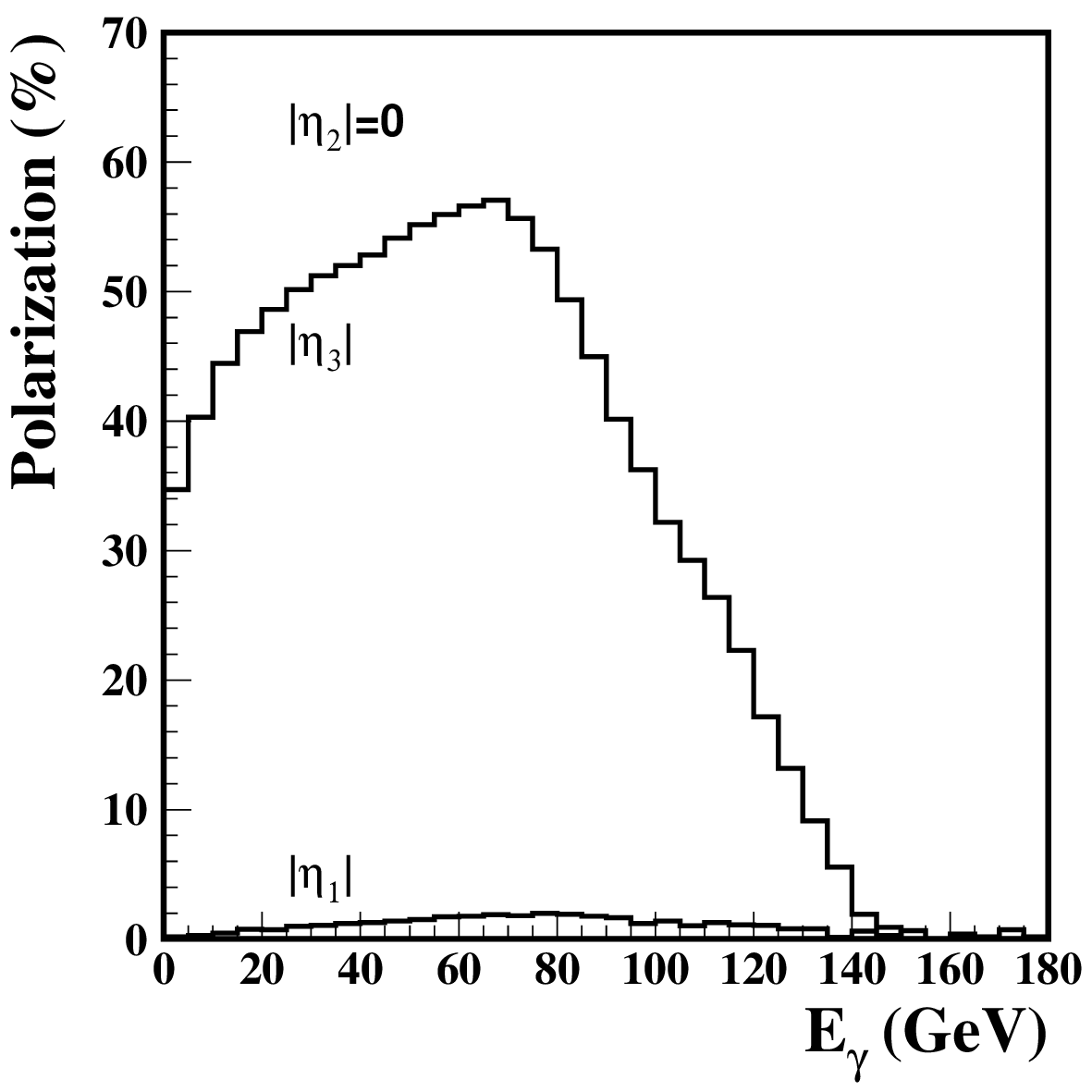}
\caption{\label{fig:pol-predict}Theoretical calculation of the polarization for 178 GeV photon beam for the radiator conditions mentioned in the text.}
\end{figure}

For an 178\,GeV electron beam making an angle of 5\,mrad from the $<$001$>$ 
crystallographic axis and about 70\,$\mu$rad from the (110) plane, the 
resulting photon beam polarization spectrum was predicted to yield maximum 
polarization of about 55\% in the vicinity of 70\,GeV, as seen in 
Fig.~\ref{fig:pol-predict}.

For this choice of crystal orientation the incidence angles of electrons and 
photons to the crystal plane become comparable with the radiation and pair 
production characteristic angle $\theta_v$. In case of the $(110)$ plane of 
the silicon crystal, we find $\theta_v$=42 $\mu$rad. In fact part of the 
incident electron beam penetrates the crystal with angles both less and 
greater than $\theta_v$, because of the angular divergence of the electron 
beam. 

In the simulations presented here, a Monte Carlo approach 
was used to model the divergence of the electron and photon beams, and the 
relevant theories (CB and CPP or the quasi classical theory) are selected as 
appropriate for accurate and fast calculation. The details may be found in 
reference~\cite{apyan-mc} and references therein. 
Briefly, the random trajectory method is deployed, 
where each particle is described by its history in propagation 
through the aligned crystals. Each particle 
history $i$ is represented by the array $\mathbf S^i_j$ 
denoting the state of particle before $j_{th}$ interaction~\cite{morin}:

\begin{equation}
\mathbf{S}^i_j = \big( \mathbf{r}^i_j , \mathbf{\Omega}^i_j ,
\mbox{\boldmath$\eta$}^i_j , E^i_j , q^i_j \big .)
\label{eq:state}
\end{equation}
where $ \mathbf{r}^i_j , \mathbf{\Omega}^i_j , \mbox{\boldmath$\eta$}^i_j ,
E^i_j , q^i_j $ represent the electron or photon position, direction,
polarization, energy and charge before each interaction acts, respectively. The
simulation code calculates the new state of particles after each interaction acts. 
A history is terminated when the particle energy drops below a low energy cut-off,
or when the particle moves outside the target. All successive interactions of 
electrons and photons with atoms are simulated, such as coherent and incoherent
bremsstrahlung and pair production. The Monte Carlo code tracks all 
of the charged particles and photons generated through the aligned crystal by
taking into account the parameters of the incoming beam, multiple scattering,
energy loss, emission angles, transverse dimension of the propagating beams, 
and the linear polarization of the photons produced. 
The corresponding energy losses, polarization  
and scattering angles are determined from the appropriate differential 
cross-sections of CB and CPP. It will be shown later that 
this approach has lead to a very good agreement between the theoretical 
predictions and the data.

\subsection{\label{subsec:SoS-prod} Enhanced production of SOS photons}

The character of the radiation, including its linear polarization, is changed 
when the direction of the electron (i) has a small angle with a 
crystallographic axis and (ii) is parallel with the plane that is formed by 
the atomic strings along the chosen axes. This is the so-called SOS 
orientation. It produces a harder photon spectrum than the CB case because the 
coherent radiation arises from successive scattering off the axial potential, 
which is deeper than the planar potential. The radiation phenomena in single 
crystals aligned in the SOS mode have been under active theoretical investigation 
since the discovery of two distinct 
photon peaks, one in the low energy region and one in the high energy region 
of the radiated energy spectrum for about 150 GeV electrons traversing a 
diamond crystal~\cite{new-effect}. It was established that the hard photon 
peak was a single photon peak \cite{kirsebom01}. However, the radiated photons 
were generally emitted with significant multiplicity in such a way that a 
hard photon would be accompanied by a few low energy photons. It will be seen 
later that two different mechanisms are responsible for the soft and the hard 
photons. In the former case, it is planar channeling (PC) radiation, while in 
the latter case, it is SOS radiation. An additional intriguing feature of SOS 
radiation at these energies ($E_{\gamma} \approx 120$ GeV) is that it occurs 
at the onset of a regime where strong field effects need to be taken into 
account. These fields are characterised by the parameter 
$\chi = \gamma \mathcal{E} / \mathcal{E}_0$ where $\gamma \mathcal{E}$ is the 
boosted crystal field in the electron frame and $\mathcal{E}_0$ is the 
Schwinger field. This is defined as the field which separates a virtual pair 
by the electron Compton wavelength, 
$\mathcal{E}_0 = m/e\lambda_c$~\cite{schwinger1, schwinger2, schwinger3}. 
The quantum suppression of radiation expected under these 
conditions~\cite{baier, sorensen, yuri1} was 
evidenced~\cite{kirsebom01, spin-flip, ulrik}, as well as 
other effects~\cite{Baurichter, Kirsebom-NIM119}. Other situations where such 
conditions have been achieved are terawatt laser fields and above barrier very 
heavy ion collisions.

The issue of the polarization of SOS radiation also came into question. Early 
experiments with electron beams of up to 10 GeV in single crystals showed a 
smaller linear polarization of the more intense radiation in the SOS 
orientation than in the PE orientation (see~\cite{saenz} and references 
therein). The first measurements of linear polarization for high energy 
photons ($E_{\gamma} \approx 50-150$ GeV) were consistent with a high degree 
of linear polarization of the radiated photons~\cite{kirsebom99}. At this 
stage the theoretical prediction of the SOS hard photon polarization was 
unresolved. It was however clear that the photons emitted by the PC mechanism 
would be  linearly polarized. The polarimeter in this experiment recorded 
the integral polarization for a given radiated energy, which was likely to 
have a multi-photon character. This experiment therefore could not be 
considered conclusive as it did not separate the PC and SOS components and 
the extent to which pile-up from the low energy photons perturbed the high 
energy part of the total radiated energy spectrum was not resolved. These 
results therefore required more theoretical and experimental investigation.

A theory of photon emission by electrons along the SOS orientation of single 
crystals has since been developed. The theory takes into account the change 
of the effective electron mass in the fields due to the crystallographic 
planes and the crossing of the atomic strings~\cite{bks}. Those authors show 
that the SOS specific potential affects the high energy photon emission and 
also gives an additional contribution in the low energy region of the 
spectrum. In references~\cite{simon,strakh3} the linear polarization of the 
emitted photons was derived and analyzed for different beam energies and 
crystal orientations. The predicted linear polarization of hard photons 
produced using the SOS orientation of the crystal is small compared to  the 
comparable case using the PE orientation of the crystal. On the other hand, 
the additional soft photons produced with SOS orientation of the crystal are 
predicted to  exhibit a high degree of polarization.

The peak energy of the SOS photons, $E_\gamma$, is determined from the same 
condition as for CP photons (equation~\ref{eq:E-peak}). However, for the SOS 
orientation, $d$ is the spacing between the axes (strings) forming the planes, 
and $\Theta=\theta$, the electron incident angle with respect to the axis.

With the appropriate choice of $\theta$ the intensity of the SOS radiation may 
exceed the Bethe-Heitler radiation (incoherent bremsstrahlung\,(ICB)) by an 
order of magnitude.

When a thin silicon crystal is used with an electron beam of energy 
$E_0 = 178$ GeV incident along the SOS orientation, within the $(110)$ plane 
and with an angle of $\theta=0.3$ mrad to the $<100>$ axis, the hard photon 
peak position is expected at $E_{\gamma}=129$ GeV.

In the current experiment, a 1.5~cm thick silicon crystal was used in the 
SOS orientation as mentioned above. Under this condition the radiation is 
expected to be enhanced by about a factor 20 with respect to the ICB for a 
randomly oriented crystalline Si target of the same thickness.

The radiation spectrum with the crystal aligned in SOS orientation has in 
addition to the CB radiation a strong component at a low energy which is 
characteristic of planar channelling (PC) radiation. As the electron direction 
lines up with a crystallographic plane in the SOS orientation, the planar 
channelling condition is fulfilled. For channelling radiation the coherence 
length is much longer than the interatomic distances and the long range 
motion, characteristic of planar channelled electrons, becomes dominant over 
short range variations with the emission of low energy photons. Theoretical 
calculations~\cite{strakh2,armen2} predict a more intense soft photon 
contribution (PC) with a high degree of linear polarization of up to 70\%.

\begin{figure}[htbp]
\includegraphics[width=0.45\textwidth]{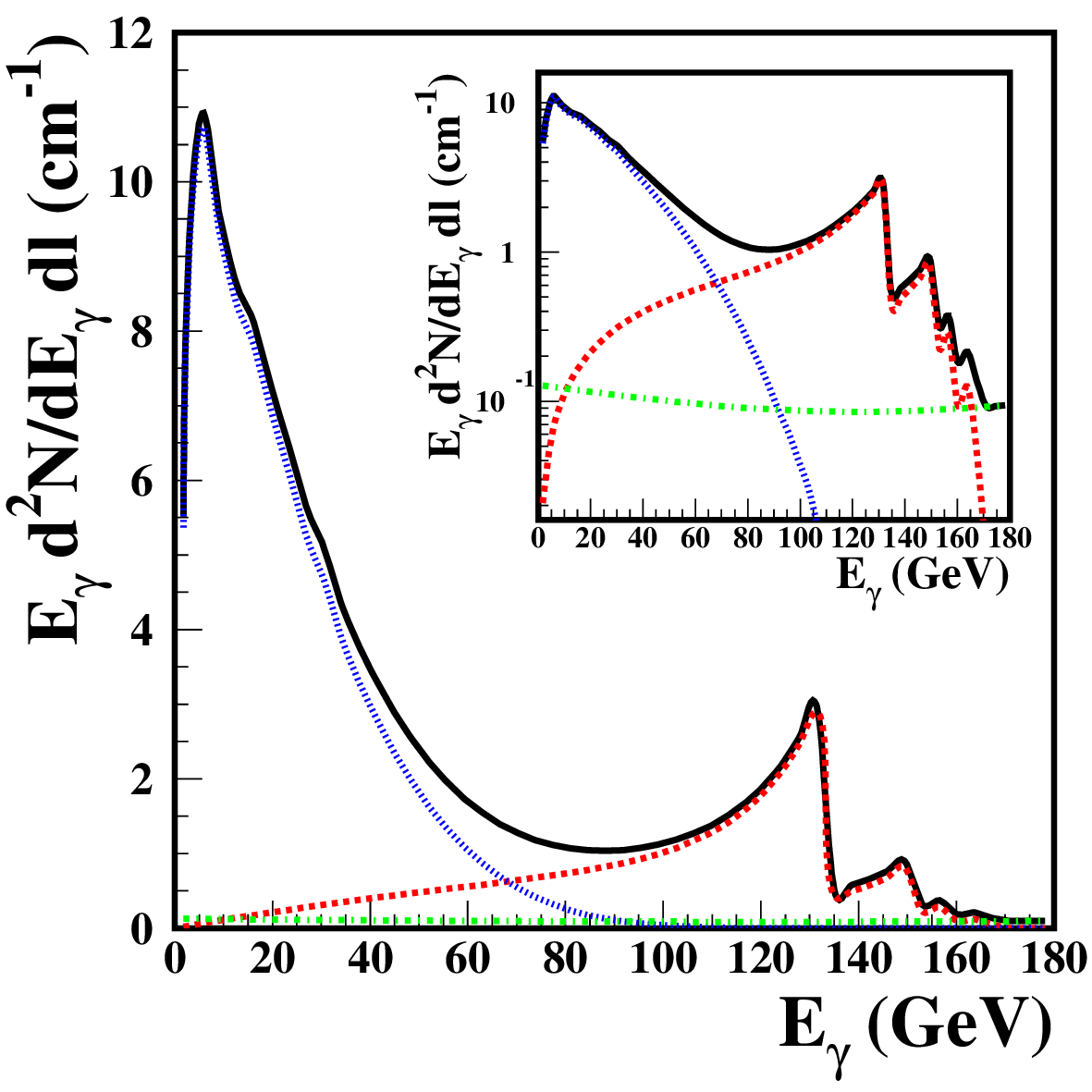}
\caption{\label{F:Strak-1b}Photon power yield, $E_\gamma d^2N/dE_\gamma dl$,
per unit of thickness for a thin silicon crystal in the SOS orientation. The
solid curve represents the total of the contributions from
ICB~(green dash-dotted), PC~(blue dotted) and SOS~(red dashed) radiation. The
insert is a logarithmic representation and shows the flat incoherent
contribution and the enhancement with a factor of about 20 for SOS radiation
at 129\,GeV.}
\end{figure}

The calculations of the enhancements of both the low energy and the high energy 
components of the radiation emission for the SOS orientation under conditions 
applicable to this experiment are presented in Fig.~\ref{F:Strak-1b}. 
Where $E_0 =178$\,GeV electron beam incidences the $(110)$ plane and at an 
angle of $\theta=0.3$\,mrad to the $\langle 100 \rangle$ axis. At low energy 
the PC radiation dominates and at high energies the SOS radiation peaks.

\section{\label{sec:beam-optics}Beam optic elements for very high energy 
photon beams}

In this work the polarization observables for high energy photons produced 
either by CB or SOS radiation are determined using the birefringence phenomena 
in CPP on aligned crystals. The study of the conversion of linear to circular 
polarization is based on the same birefringent effect in CPP. The experiments 
and their theoretical simulation therefore represent simultaneously a test of 
the theoretical understanding as well as development of new beam-optic 
elements based on crystal techniques.

\subsection{\label{subsec:Xtal-pol}Birefringence in CPP and crystal polarimetry}

Historically, the pair conversion in single crystals was proposed, and later 
successfully used in the 1960s as a method to measure linear polarization for 
photons in the 1-6 GeV range~\cite{barbiellini}. It was predicted 
theoretically and later verified experimentally~\cite{expver} that the pair 
production cross section and the sensitivity to photon polarization increases 
with increasing energy. Therefore, at sufficiently high photon energies, a 
new polarization technique based on this effect can be constructed, which 
will become competitive to other techniques, such as pair production in 
amorphous media and photo nuclear methods.

In the first part of the experiment, the cross section for CPP by polarized 
photons incident on the aligned "analyzer" crystals (germanium and diamond) 
was measured, for different carefully selected crystallographic orientations. 
This process can be effectively viewed as the imaginary part of the refractive 
index, as it leads to an attenuation of the photon beam. It constitutes a 
birefringence phenomenon, as the imaginary part of the refractive index will 
differ as a function of the angle between the plane of polarization of the 
photon beam and a specific crystallographic orientation of the "analyzer" 
crystal. A polarimeter was constructed by measuring the energy dependent 
asymmetry with respect to the two most distinct orientations of the analyzer 
crystal for pair production.

The theoretical comparison to the data could validate the calculation of the 
energy dependence of the cross section and the polarization of photons 
produced by coherent bremsstrahlung as well as the calculation of coherent 
pair production for polarized photons incident on crystals of different 
crystallographic orientations.

The photon polarization is expressed using the Stoke's parameterization with 
the Landau convention, where the total elliptical polarization is decomposed 
into two independent linear components and a circular component. Referred to 
our geometry the parameter $\eta_1$ describes the linear polarization of the 
beam polarized in the direction of $45^{\circ}$ to the reaction plane of the 
radiator, while the parameter $\eta_3$ describes the linear polarization in 
the direction parallel or perpendicular to the reaction plane of the radiator. 
The parameter $\eta_2$ describes the circular polarization. The total 
polarization is then written:

\begin{widetext}
\begin{equation}
P_{\hbox {linear}}=\sqrt{\eta _{1}^{2}+\eta _{3}^{2}},
\quad \; P_{\hbox {circular}}=\sqrt{\eta _{2}^{2}},
\quad \; P_{\hbox {total}}=\sqrt{P_{\hbox {linear}}^{2}+P_{\hbox
{circular}}^{2}} \quad .
\label{eq:pol-def}
\end{equation}
\end{widetext}

The radiator angular settings were chosen to have the total linear 
polarization from CB radiation  purely along $\eta _{3}$. Two distinct 
measurements were made, one to show that the $\eta _{1}$ component of the 
polarization was consistent with zero and another to find the expected 
$\eta _{3}$ component of polarization as shown in Fig.~\ref{fig:pol-predict}.
The Monte Carlo calculations used to obtain this prediction took into account the 
divergence of the electron beam ($48\,\mu$rad horizontally and 33\,$\mu$rad 
vertically) and the 1\% uncertainty in its 178~GeV energy. To optimise the 
processing time of the Monte Carlo simulation, minimum energy cuts of 5~GeV for the 
electrons and 500 MeV for the photons were applied. We were, therefore, able 
to predict both the total radiated energy spectrum and the energy spectrum of 
individual photons.

The polarization dependence of the pair production cross section and the 
birefringent properties of crystals are key elements of the photon 
polarization measurement. The imaginary parts of the refraction indices are 
related to the pair production cross section. This cross section is 
sensitive  to the relative angle between a crystal plane of a specific 
symmetry and the plane of linear polarization of the incident photon. In 
essence, the two orthogonal directions where these two planes are either 
parallel or perpendicular to each other yield the greatest difference in pair 
production cross section.

Thus, the dependence of the CPP cross section on the linear polarization of 
the photon beam makes an oriented single crystal suitable as an efficient 
polarimeter for high energy photons. The existence of a strong anisotropy for 
the production of the e$^+$e$^-$ pairs during their formation is the reason 
for the polarization dependent CPP cross section of photons passing through 
oriented crystals. This means that perfect alignment along a crystallographic 
axis is not an efficient analyzer orientation due to the approximate 
cylindrical symmetry of the crystal around atomic strings. However, for small 
angles of the photon beam with respect to the crystallographic symmetry 
directions the conditions for the formation of the e$^+$e$^-$ pairs prove to 
be very anisotropic. As it turns out, the orientations with the highest 
analyzing power are those where the e$^+$e$^-$ pair formation zone is not only 
highly anisotropic but also inhomogeneous with maximal fluctuations of the 
crystal potential along the electron path. At the crystallographic axes the 
potential is largest and so are the fluctuations. These conditions are related 
to the ones of the SOS orientation: (i) a small angle to a crystallographic 
axis to enhance the pair production process by the large fluctuations and 
(ii) a smaller angle to the crystallographic plane to have a long but still 
anisotropic formation zone for CPP.

We therefore studied the pairs created in a second aligned crystal, called 
the {\em analyzer} crystal. In this study, the experimentally relevant 
quantity is the asymmetry, $A$, between the pair production cross sections, 
$\sigma$, of parallel and perpendicular polarized photons, where the 
polarization direction is measured with respect to the $<$110$>$ 
crystallographic plane of the analyzer crystal. This asymmetry is related to 
the linear photon polarization, $P_{\rm l}$, through the equation

\begin{equation}
A  \equiv   \frac{\sigma (\gamma _{\perp }\rightarrow e^{+}e^{-})-\sigma
(\gamma _{\parallel }\rightarrow e^{+}e^{-})}{\sigma (\gamma _{\perp
}\rightarrow e^{+}e^{-})+\sigma
(\gamma _{\parallel }\rightarrow e^{+}e^{-})}
= R \times P_{\rm l}.
\label{eq:asy-def}
\end{equation}

Here $R$ is the so called ``analyzing power'' of the second crystal. The 
analyzing power is in fact the asymmetry expected for a 100\% linearly 
polarized photon beam. It will be seen that for the conditions of this 
experiment, and using the theory described, this quantity can be reliably 
computed using Monte Carlo simulations. In this polarimetry method, the 
crystal with the highest possible analyzing power is preferred in order to 
achieve a fast determination of the photon polarization.

\begin{figure}[ht]
\includegraphics[width=0.45\textwidth]{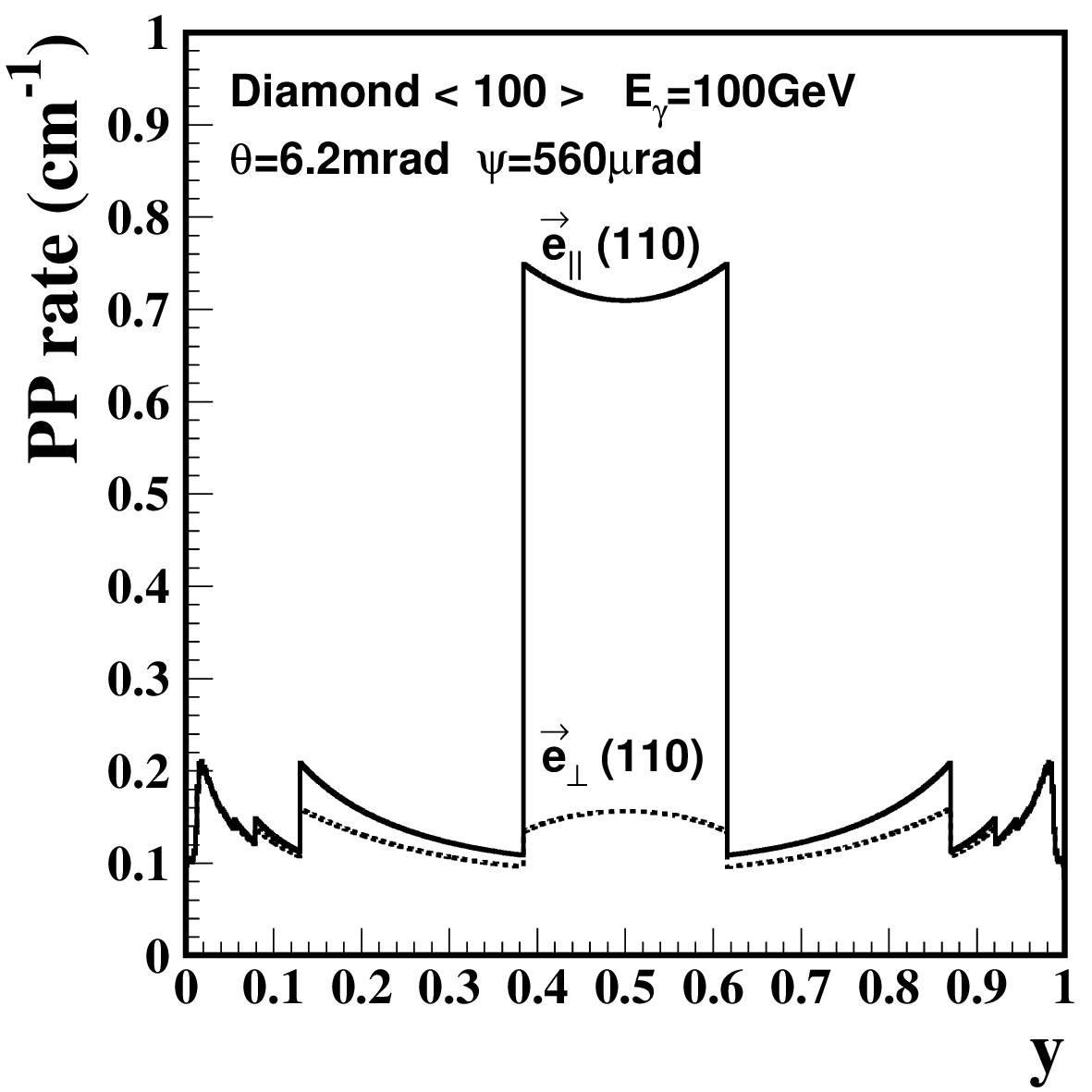}
\caption{\label{fig:ycut} Pair production rate vs the pair asymmetry, $y$, 
as defined in the text.}
\end{figure}

If one defines a parameter to measure the pair energy asymmetry as the 
ratio of the energy of one of the pairs, $E^-$, to the 
energy of the incoming photon, $E_\gamma$, as

\begin{equation}
y\equiv E^-/E_\gamma \quad ,
\end{equation}
then one may calculate the dependence of the pair production rate on this 
ratio, $y$, as shown  in Fig.~\ref{fig:ycut}
(zero asymmetry corresponds to $y = \frac{1}{2}$).

\begin{figure*}[htbp]
\includegraphics[scale=0.48]{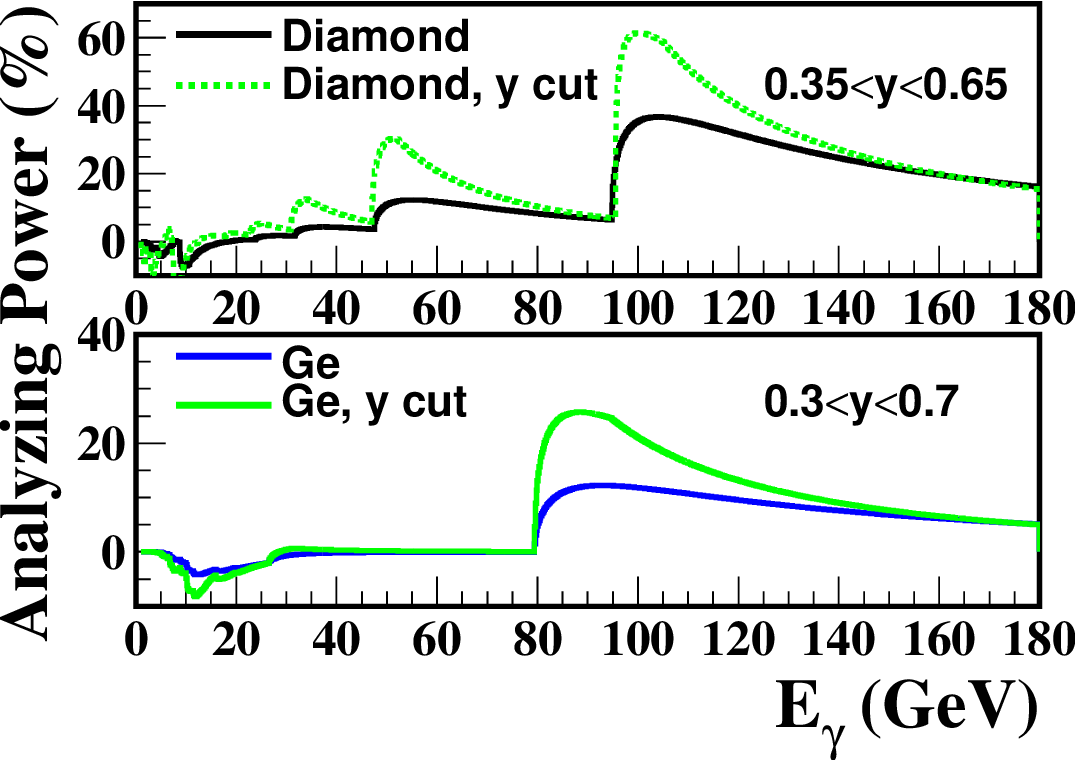}
\includegraphics[scale=0.48]{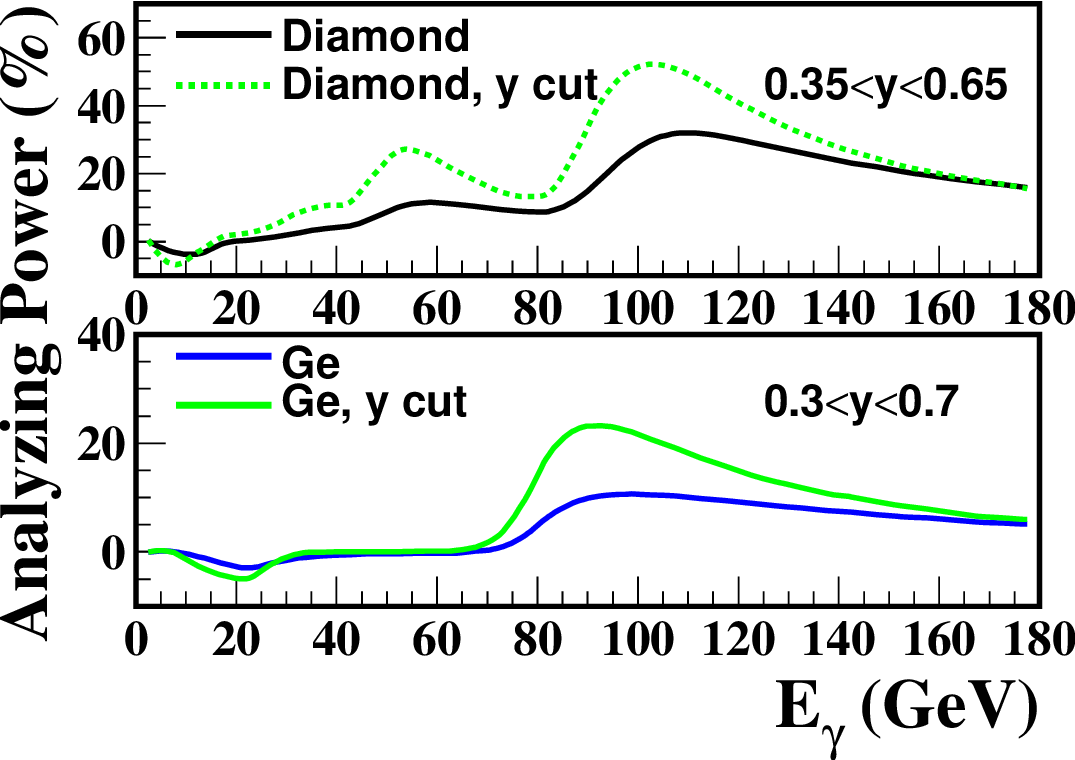}
\caption{\label{fig:ap} Analyzing power of different single crystals, for an
ideal $e^-$ beam without any angular divergences (left) and for the actual $e^-$
beam conditions (right).}
\end{figure*}

By comparing the rates for the photon polarization parallel (solid line) and 
perpendicular (dashed line) to the crystallographic plane, we observe that the 
largest difference arises for \mbox{0.4 $\leq  y \leq$ 0.6}. Therefore the 
pair production asymmetry may be maximised by selecting the subset of events 
where the $e^{+}e^{-}$ pairs have similar energies. This method of choosing 
the pairs to enhance the analyzing power is called the ``quasi-symmetrical 
pair selection method''~\cite{ycut}. As a result of such a cut, although the 
total number of events decreases, the relative statistical error diminishes 
since it is inversely correlated with the measured asymmetry. If the 
efficiencies of the pair events and beam intensity normalisation events are 
assumed to be the same, then the cross section measurement in 
equation~(\ref{eq:asy-def}) reduces to counting these events separately. 
Denoting the number of pairs produced in perpendicular and parallel cases by 
$p_{1}$ and $p_{2}$, and the number of the normalisation events in each case 
by $n_{1}$ and $n_{2}$, respectively, the measured asymmetry can be written as:

\begin{equation}
A=\frac{p_1/n_1 - p_2/n_2}{p_1/n_1 + p_2/n_2},
\label{eq:asy-meas}
\end{equation}
where $p$ and $n$ are acquired simultaneously and therefore are correlated.

\subsection{\label{subsubsec:crystals}Germanium and diamond analyzer crystals}

The first analyzer crystal used in the experiment was a germanium (Ge) 
single crystal disk with a diameter of 3\,cm and a thickness of 1.0\,mm. The 
selected orientation  with respect to the incident photon beam represented a 
polar angle of 3.0\,mrad measured from the $<$110$>$ axis and an azimuthal 
angle corresponding to incidence exactly on the (1$\bar{1}$0) plane. This 
configuration gave an analyzing power peaking at 90~GeV, as can be seen in 
Fig.~\ref{fig:ap}. From the same figure one can also see that the 
quasi-symmetrical pair selection method delivers almost twice the analyzing 
power. The same single Ge crystal had also been used in the a previous 
experiment, as reported in~\cite{Na43ge} therefore the pair production 
properties of this thickness of germanium crystal are well known.

The second analyzer for this experiment was a multi-tile synthetic diamond 
crystal target with an incident photon beam orientation with respect to the 
crystal of 6.2\,mrad from the $<$100$>$ axis and 560\,$\mu$rad from  the 
(110) plane.

\begin{figure}[htbp]
\includegraphics[width=0.45\textwidth]{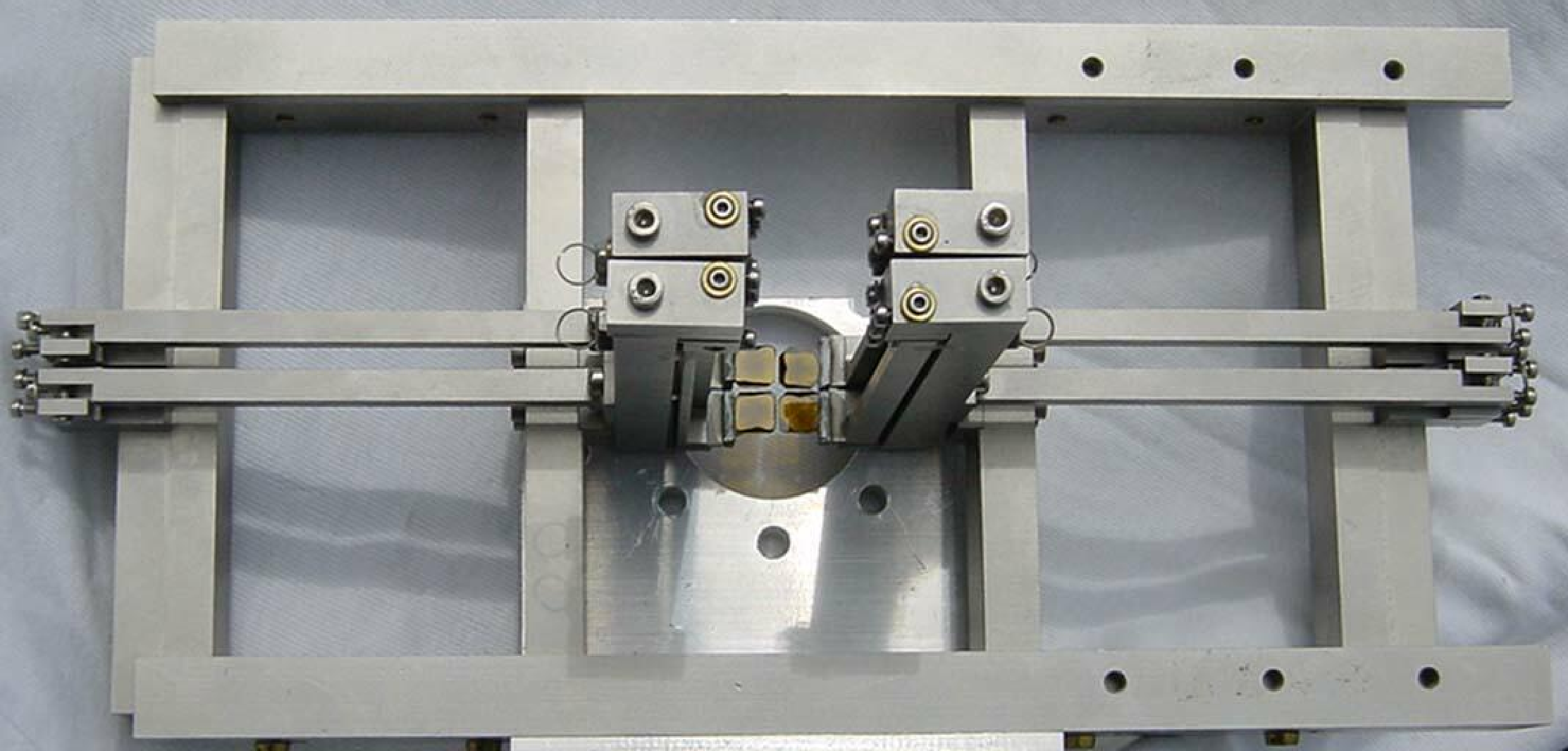}
\caption{\label{fig:diamonds} The diamond analyzer target consists of
          synthetic diamond tiles and the aluminium holder frame.}
\end{figure}

The major advantage of using diamond in the analyzer role are its
high pair yield, high analyzing power (see Fig.~\ref{fig:ap}) and radiation 
hardness. The photon beam dimensions of this experiment implied that one would need a 
diamond with an area of about 20mm$\times$20mm. A crystal thickness of 4~mm 
was a fair compromise between requirements of the Figure of Merit (FOM) for a 
diamond analyzer and the costs of the material. These requirements were 
realised by developing a composite target comprising of four synthetic 
type Ib diamonds \cite{pap3,pap4,pap5a} of 
dimensions 8$\times$8$\times$4~mm$^3$ arranged in a square lattice 
as seen in Fig.~\ref{fig:diamonds}.

Once beam was available, the fine alignment was performed (and indeed regularly 
controlled during the experiment). A narrow electron beam  was directed onto 
the crystal, and data was collected using the minimum bias trigger 
(see section IV A). A scan of 
the incident angular phase space between the beam and the crystal was 
performed by programming the motion of the crystal mounted in the goniometers. 
The crystallographic axes and planes could be identified as positions in this 
phase space where the coherent enhancements (or reductions) of a radiation 
phenomenon in relation to the corresponded incoherent cross-section occurred. 
This would be observed in an appropriate detector.

The radiator crystal was therefore aligned exploiting the physics of 
bremsstrahlung from the electron beam as observed in the Lead Glass 
Calorimeter. On the other hand, the analyzer crystal was aligned by observing 
pair-production by the photon beam generated in the radiator crystal as 
observed in the multiplicity counter.

\subsection{\label{subsec:CircPol}Conversion of linear to circular polarization}

\begin{figure*}[htpb]
\includegraphics[scale=0.42]{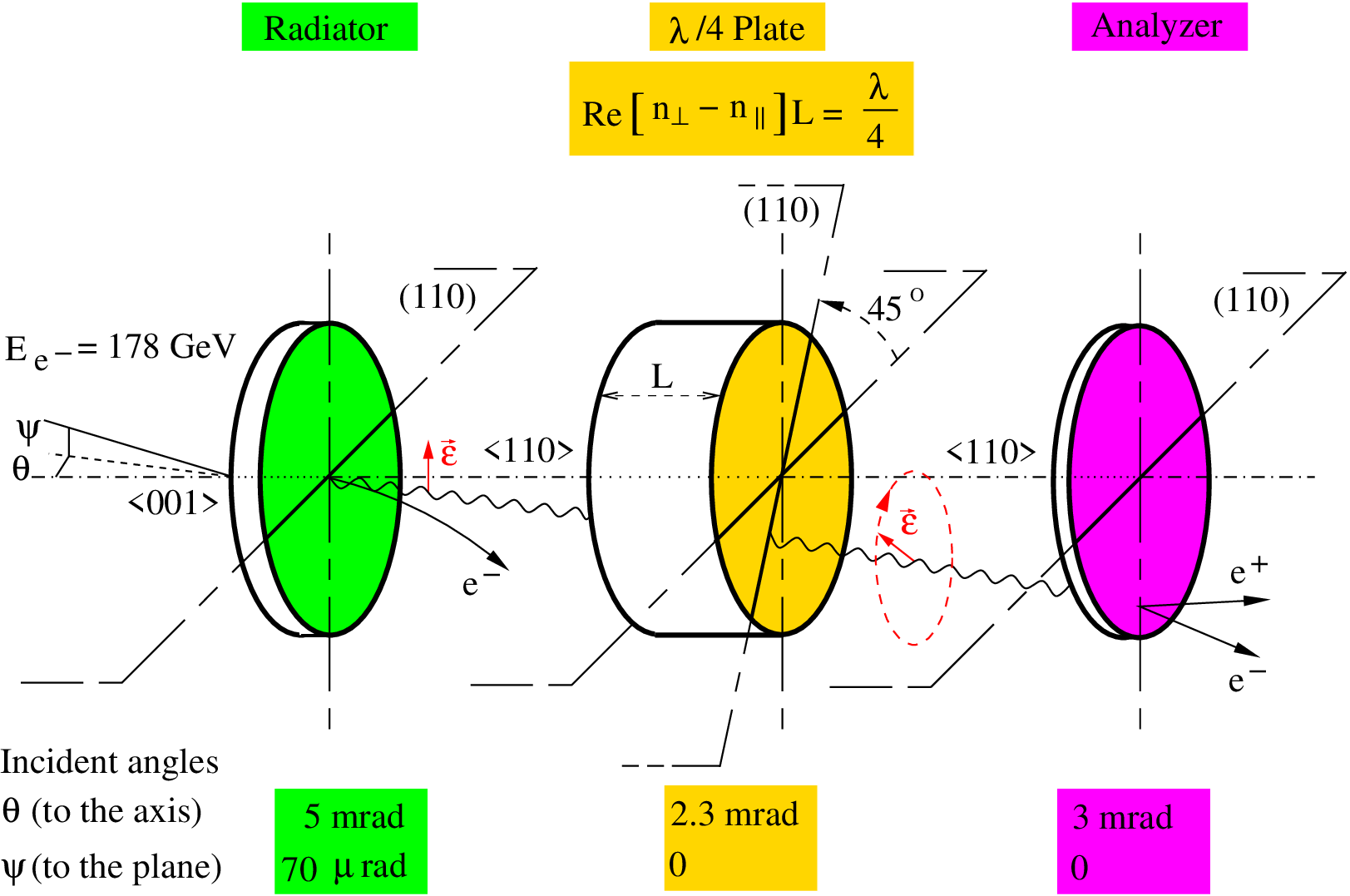}
\caption{\label{F:setup1} Three crystal scheme.}
\end{figure*}

The second section of the experiment tested the feasibility of producing 
circularly polarized photon beams in proton accelerators using the extracted 
unpolarized high energy electron beams (for example, with energies of up to 
250\,GeV (CERN) and 125\,GeV (FNAL)~\cite{proposal,na59-pap1}). These unpolarized 
electron beams can produce linearly polarized photons via CB radiation in an 
aligned single crystal. One can transform the initial linear polarization into 
circular polarization by using the birefringent properties of aligned 
crystals. The above mentioned method was first proposed by Cabibbo and 
collaborators in the 1960's~\cite{cabibbo1}, and later the numerical 
calculations were done in terms of CPP theory~\cite{termisha} to obtain the 
optimal thicknesses for various cubic crystals. To perform the experimental 
investigation, a consecutive arrangement of three aligned single crystals was 
used. The first crystal acted as a radiator  to produce a linearly polarized 
photon beam, the second crystal acted as a quarter wave plate to convert the 
linear polarization into circular polarization, and the last crystal acted as 
an analyzer to measure  the change in the linear polarization of the photon 
beam. The three-crystal scheme used is shown in Fig.~\ref{F:setup1}.

The linearly polarized photon beam was produced by CB radiation from electrons 
in an aligned Si\,$<$100$>$ single crystal (radiator), as already described above 
(section \ref{subsec:CB-prod}). For the conversion of the linear polarization 
into circular polarization an  aligned Si\,$<$110$>$ crystal was used 
(quarter wave plate). Finally the resulting polarisation of photon beam 
was measured in an aligned Ge$<$110$>$ crystal (analyzer), also as discussed 
above (section~\ref{subsec:Xtal-pol}).

When a high energy photon beam propagates through a medium, the main process by 
which the photons are absorbed is e$^+$e$^-$ pair production. When photons 
propagate through an aligned crystal at small incident angles with respect to 
a crystal axis and/or a crystal plane, a coherent enhancement of the PP is 
manifested (CPP). The cross section for the CPP process depends on the 
direction of the linear polarization of the photon beam with respect to the 
crystal axis and to the photon momentum (reaction plane) as shown in 
Fig.~\ref{F:setup1}. Generally speaking, one can represent the linear 
polarization of the photon beam as a superposition of two beams with 
polarization directions parallel and perpendicular to the reaction plane 
containing the photon momentum {\boldmath $k$} and the crystallographic axis. 
In this case, the photon polarization vector {\boldmath $e$} will be the 
combination of two unit vectors, {\boldmath $t$} and {\boldmath $y$}, parallel 
and perpendicular to the reaction plane, respectively:

\begin{equation}
\mbox{\boldmath $e$} = \mbox{\unboldmath$e_{\parallel}$} \mbox{\boldmath $t$} +
\mbox{\unboldmath$e_{\perp}$} \mbox{\boldmath $y$}.
\label{eq:no1}
\end{equation}

The components of the polarization vector before and after the crystal of
thickness, $L$, are related by a $2 \times 2$ matrix~\cite{cabibbo1,zorzi}:

\begin{displaymath}
\left( \begin{array}{c}
e_{\parallel}(L)\\
e_{\perp}(L)
\end{array} \right)
=
\left( \begin{array}{cc}
      \exp [in_{\parallel} E_\gamma L] & 0 \\
      0 & \exp [in_{\perp} E_\gamma L]
\end{array} \right)
\left( \begin{array}{cc}
      e_{\parallel}(0) \\
      e_{\perp}(0)
\end{array} \right)
\end{displaymath}
where $E_\gamma$={\boldmath $|k|$} ($\hbar=c=1$) is the energy of the incident 
photon and the $n_\parallel$ and $n_\perp$ are complex quantities analogous to 
the index of refraction. The imaginary part of the index of refraction is 
connected with the photon absorption cross section, while the real part can 
be derived from the imaginary part using dispersion relations~\cite{cabibbo1}. 
The crystal can act as a {\it quarter wave plate}, if the real part of the 
relative phases of the two components of the waves parallel and perpendicular 
to the reaction plane is changed by $\pi/2$ after transmission of the photon. 
Thus the crystal will be able to transform the linear polarization of the 
photon beam into circular polarization at the matching thickness:

\begin{equation}
L = \frac{2}{\pi}\frac{1}{E_\gamma\Re(n_{\perp} - n_{\parallel})}.
\label{eq:no2}
\end{equation}

The polarization is expressed again as in section \ref{subsec:Xtal-pol}. The 
photon beam intensity and Stokes parameters after the quarter wave plate with 
the thickness $L$ can be derived from the following formulae~\cite{baier}:

\begin{eqnarray}
N(L) &=& N(0)[\cosh aL+\eta _{1}\sinh aL]\exp (-WL),
\nonumber \\
\eta_1 (L) &=& \frac{\sinh aL + \eta_1 (0)\cosh aL}
              {\cosh aL +\eta_1 (0)\sinh aL},
\nonumber \\
\eta_2 (L) &=& -\frac{ \eta_3 (0)\sin bL - \eta_2 (0)\cos bL}
              {\cosh aL + \eta_1 (0)\sinh aL},
\nonumber \\
\eta_3 (L) &=& -\frac{\eta_3 (0)\cos bL + \eta_2 (0)\sin bL}
                {\cosh aL + \eta_1 (0)\sinh aL},
\label{eq:no4}
\end{eqnarray}
with
\begin{eqnarray}
a &=& E_\gamma\Im(n_{\perp} - n_{\parallel}) = \frac{1}{2}
(W_{\parallel}-W_{\perp}),
\nonumber \\
b &=& E_\gamma\Re(n_{\perp} - n_{\parallel}),
\quad W = \frac{1}{2}(W_{\parallel}+W_{\perp}),
\label{eq:no5}
\end{eqnarray}
where $W_\parallel$ and $W_\perp$ are the pair production probabilities
per unit path length for photons polarized parallel or perpendicular to
the reaction plane, respectively.

As follows from equation~(\ref{eq:no4}), the component of the linear 
polarization in the direction of $45^{\circ}$ to the reaction plane of the 
quarter wave crystal is transformed into circular polarization~\cite{baier}. 
Therefore the {\it quarter wave plate} should be rotated by $45^{\circ}$ with 
respect to the polar plane of the photon beam to have the optimal 
transformation of the polarization. In this case the linear polarization 
component $\eta_3$, which was defined as the one parallel or perpendicular to 
the reaction plane of the radiator, represents a component of the linear 
polarization in the direction of $45^{\circ}$ to the reaction plane of the 
quarter wave plate equation~(\ref{eq:no4}).

As follows from equation~(\ref{eq:no4}), the total polarization of the photon 
beam before and after the quarter wave crystal are connected by the relation:

\begin{equation}
P_{total}^2(L) = 1 + \frac{P_{total}^2(0)-1} {(\cosh aL +
\eta_1 (0)\sinh aL)^2}.
\label{eq:no6}
\end{equation}

There is conservation of polarization if the incident photon beam is 
completely polarized. In a real experiment, the incident photon beam is not 
completely polarized, and one must seek an alternative conserved quantity. 
Further study of equation~(\ref{eq:no4}) reveals that the quantity

\begin{equation}
K\equiv\frac{\eta_2^2(\ell) + \eta_3^2(\ell)}{1- \eta_1^2(\ell)}
\label{eq:no7}
\end{equation}
is constant and conserved when a photon  beam penetrates the quarter wave
plate crystal~\cite{kononets}. This relation holds for any penetration length, 
$\ell$, between $0 \leq \ell \leq L$ except in the case when 
$\eta_2(0)=\eta_3(0) \equiv 0$ and $\eta_1(0)=1$. It allows the determination 
of the  resulting circular polarization of photon beam by  measuring its 
linear polarization before and after the quarter wave plate. Taking into 
account the experimental condition, i.e. the photon beam angular divergences, 
one can note that $K$ is conserved with $\approx$5$\%$ accuracy in the 
80-110~GeV region as shown by Monte-Carlo simulations.

\begin{figure}[htbp]
\includegraphics[scale=0.21]{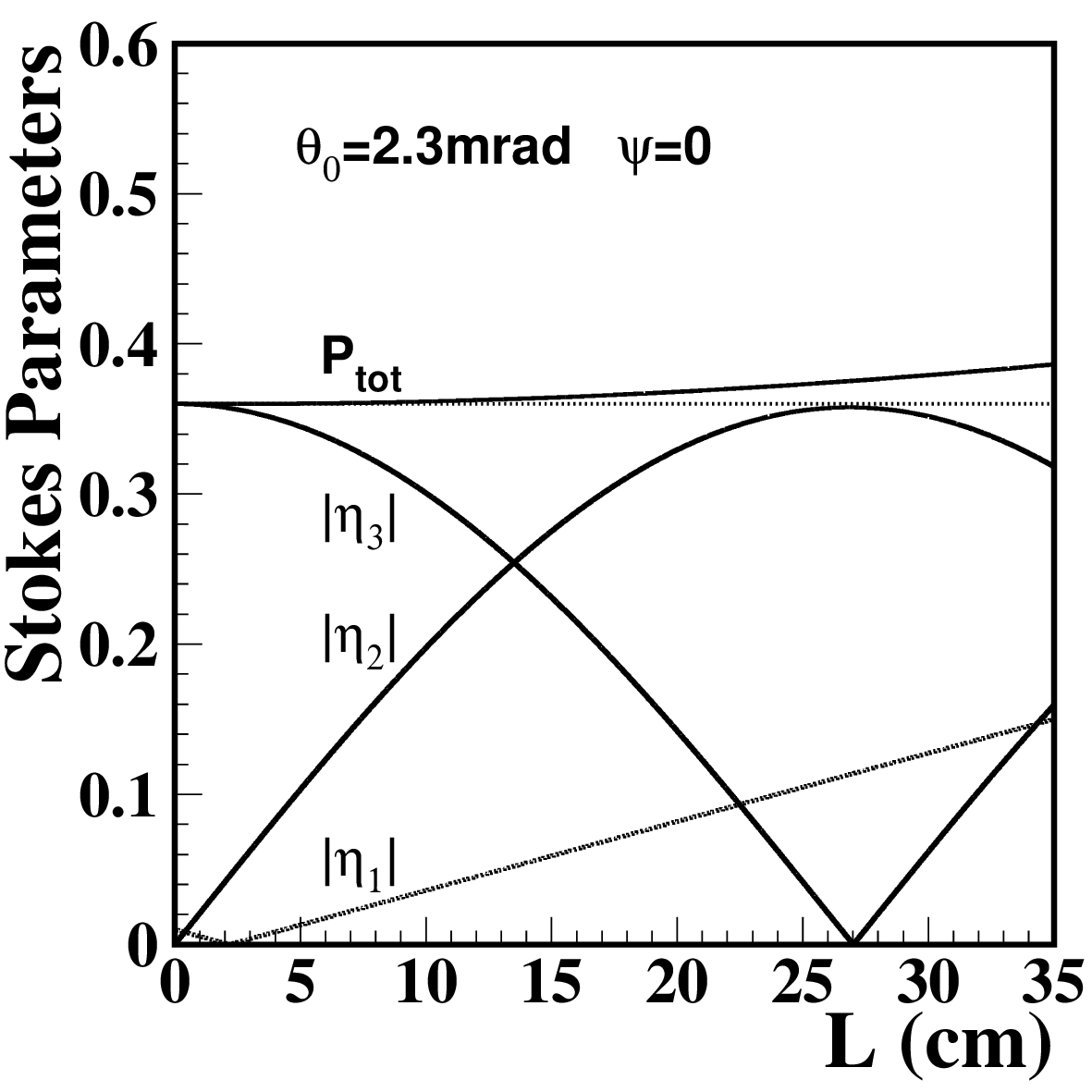}
\includegraphics[scale=0.21]{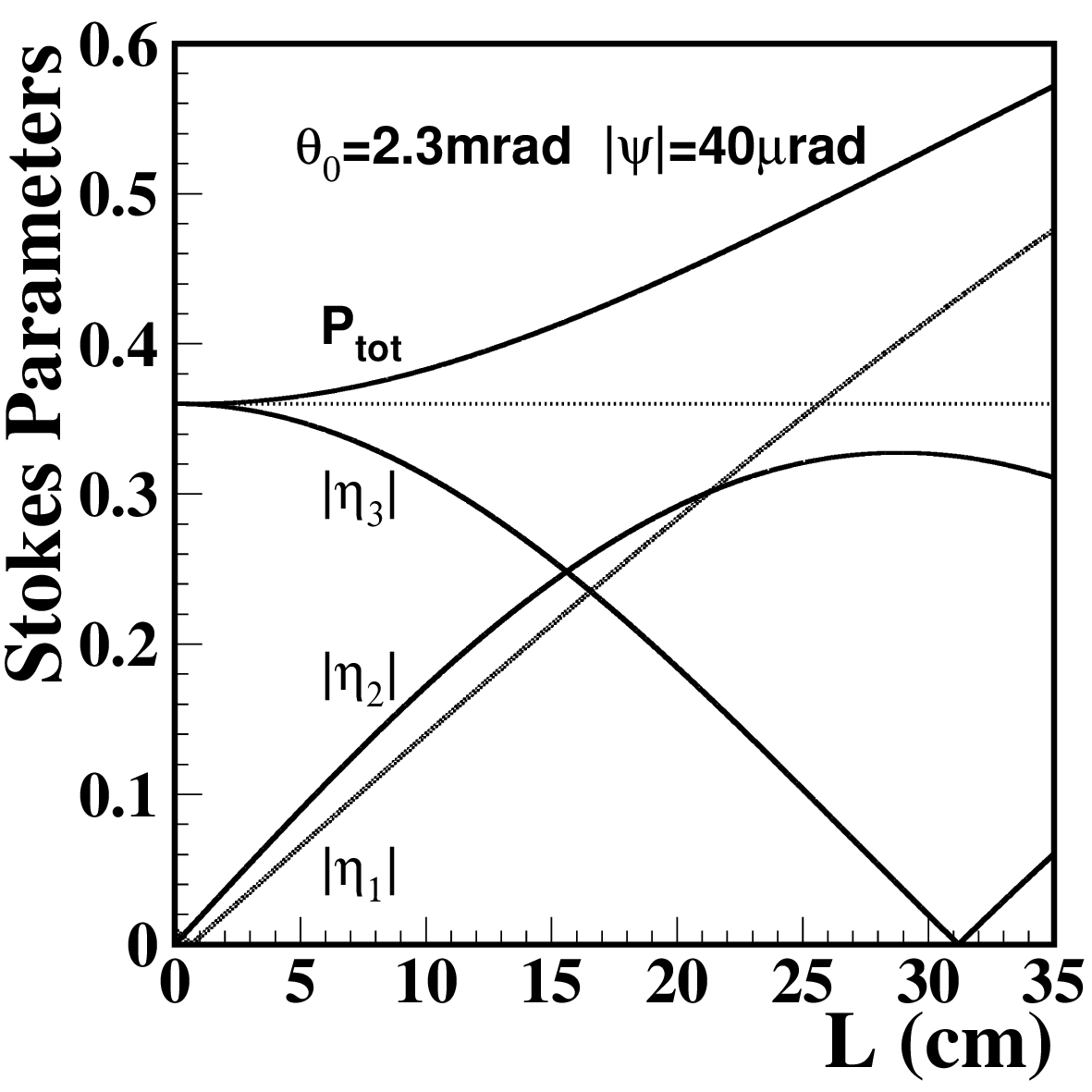}
\caption{\label{F:stokes} Absolute values of the Stokes parameters and
   the total degree of polarization for a Si crystal as a function of its
   thickness $L$, for $E_\gamma$=100\,GeV linearly polarized  photons. The
   left hand figure and the right hand figure are calculated using initial
   values for the Stokes parameters described in the text. For these
   conditions, the crystal also acts as polarizer generating a $\eta_1$
   component.}
\end{figure}

Fig.~\ref{F:stokes} shows the expected dependence of the Stokes parameters
describing the photon polarization as a function of the {\it quarter wave
plate} thickness, $\ell$, for the surviving photons from a beam of 100\,GeV.
One can see that the initial total polarization is not conserved in the case
of a partially polarized photon beam as expected from equation~(\ref{eq:no6}),
nevertheless, the relation~(\ref{eq:no6}) still holds.

These calculations were carried out assuming that the Stokes parameters before 
the {\it quarter wave plate} had the following values: $\eta_1$=0.01, 
$\eta_2$=0 and $\eta_3$=0.36. In Fig.~\ref{F:stokes} (left), the photon beam 
makes an angle $\theta_0$=2.3\,mrad with respect to the $<$110$>$ axis and 
lies in the $(110)$ plane ($\psi$=0), while in Fig.~\ref{F:stokes} (right), 
the photon beam traverses the $(110)$ plane at a small angle, 
$\psi$=$\pm$40\,$\mu$rad.

One can see the increase in the total polarization, $P_{total}$, after the 
{\it quarter wave plate} with respect to the initial total polarization (the 
straight line around 0.36). This difference comes from the fact that the 
aligned {\it quarter wave plate} can also act as a polarizer. Therefore, the 
total polarization behind the {\it quarter wave plate} can be higher than the 
initial polarization. This increase is more pronounced in the case when the 
photon momentum makes a small angle of $\psi$=40\,$\mu$rad with respect to 
the crystal plane (Fig.~\ref{F:stokes}~right). As described in 
section~\ref{subsec:exp} and as shown in Fig.~\ref{F:l4-stokes}, the final 
calculation takes into account the beam divergence, in both the horizontal 
and vertical planes.

As seen from figure~\ref{F:stokes} the Si crystal with a thickness of 
$L>$25cm has indeed acted as a quarter wave plate and generated a degree of
circular polarisation taking into account the angular 
divergence of the $\gamma$-beam. For these crystal  thicknesses where the 
$\eta_3(0)$ component of the initial linear photon beam polarization will be 
totally transformed into the final circular component $\eta_2(L)$, only a few 
percent of the  photons will survive. We defined a FOM, to find a compromise 
between the photon beam attenuation and the polarization transformation 
efficiency in~\cite{proposal}, as:

\begin{equation}
FOM = \eta_2(\ell) \sqrt {N(\ell)}.
\label{eq:no8}
\end{equation}

Here $N(\ell)$ is the statistical weight of the number of surviving photons. 
Taking into account equation~(\ref{eq:no8}), 
references~\cite{maish2,akop,strakh} 
presented theoretical predictions showing the possibility of transforming the 
linear polarization of a high energy photon beam into circular polarization in 
the 70-100\,GeV energy range. The theoretical calculations of the energy and 
the orientation dependence of the indices of refraction were performed using 
the quasi-classical operator method and CPP formulae respectively. In both 
these references, the optimum thickness for a {\it quarter wave plate} Si 
crystal was found to be 10\,cm. The relevant geometrical parameters involved 
the photon beam forming an angle of 2.29\,mrad from the axis $(110)$ and the 
photon momentum directly in the $(110)$ plane of Si single crystal, {\it i.e.} 
the angle between the photon momentum and crystal plane is $\psi$=0. For this 
choice of parameters, the fraction of surviving photons is 17-20$\%$. 

\subsection{\label{subsec:sos-rad}SOS radiation}

For the production of the much enhanced yield of the SOS photons as compared 
to the CB photons, the first radiator crystal was adjusted to appropriate 
angle settings. These were a beam angle of $\theta=0.3$ mrad to the 
$\langle 100 \rangle$ axis in the $(110)$ plane of the 1.5~cm thick Si crystal 
which is the optimal angle for a high energy SOS photon peak at 129 GeV 
(see Fig.~\ref{F:Strak-1b}) energy photon peak with a thin radiator at 125 GeV 
(see Fig.~\ref{F:Strak-1b}). As has been mentioned in 
section~\ref{subsec:SoS-prod}, the radiation probability with a thin radiator 
is expected to be 20 times larger at that energy than the Bethe-Heitler (ICB)
prediction for randomly oriented crystalline Si.

The polarization measurements were then performed in a similar way to those 
of section~\ref{subsec:Xtal-pol} for the analysis of linear polarization in CB 
radiation.

\section{\label{analysis}Experiment and Analysis}

The experiment was performed in the North Area of the CERN SPS, where 
unpolarized electron beams with energies above 100 GeV are available. We used 
a beam of 178 GeV electrons with angular divergence of $\sigma_{x'}=48$ 
$\mu$rad and $\sigma_{y'}=35$ $\mu$rad in the horizontal and vertical plane, 
respectively.
                                                                                
\subsection{\label{subsec:exp}The Setup}
\begin{figure*}[htpb]
\includegraphics[scale=0.6]{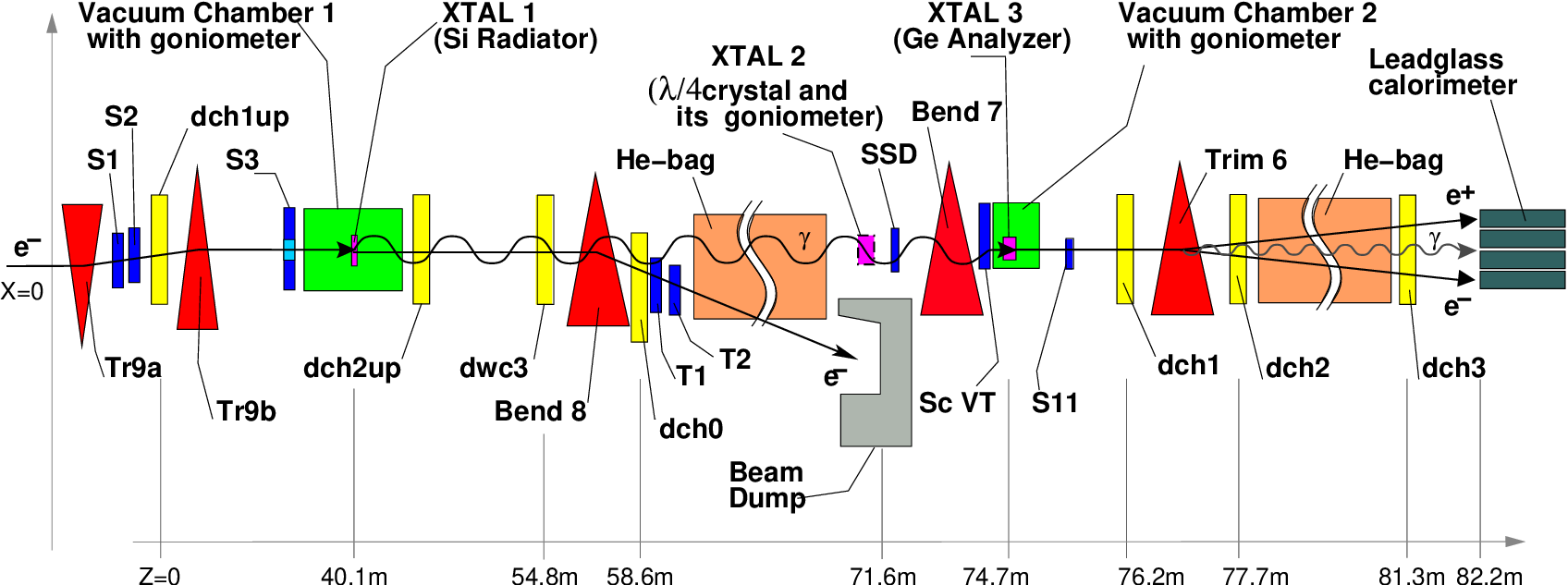}
\caption{\label{fig:exp-setup} The experimental setup.}
\end{figure*}

The experimental setup described below and shown schematically in 
Fig.~\ref{fig:exp-setup} is ideally suited for detailed studies of the photon 
radiation and pair production processes in aligned crystals.

The radiator system comprised of the 178~GeV unpolarized electron beam focused 
on the single crystal silicon radiator (XTAL1). The crystal was of cylindrical 
shape with a 2.5~cm radius and a 1.5~cm thickness. It was aligned using a 
goniometer of 2~$\mu$rad precision to obtain either CB or SOS radiation 
conditions, as required. Upstream drift chambers (dch1up-2up) allowed 
tracking of the incoming beam with an angular precision of 4\,$\mu$rad to 
define the position of incident electron in the incident angle phase space. 
The drift chambers had an active area of 15$\times$15~cm$^2$ divided into six 
cells in both horizontal and vertical planes. A double sense wire 
configuration  removed the directional hit ambiguity. The exit angles of the 
electron emerging from the radiator crystal was recorded by two tracking 
chambers (dch2up and dwc3). This allowed the measurement of electron multiple 
scattering angle inside the crystal. The dwc3 is a multi wire proportional 
chamber~\cite{dwc} with an active area of about $10\times10$ cm and a 
resolution of 200~$\mu$m.

The photon tagging system consisted of a dipole magnet (Bend8) capable of a 
maximum beam rigidity of 4.053~Tm and a special drift chamber (dch0) with no 
active horizontal cells. This constituted an upstream spectrometer which 
measured the energy of the spent electron (with an acceptance of 10-90\% of 
the incident energy). A beam dump protected the rest of the system from 
background effects arising from the spent electron.

The polarization analyzer system followed next. After passing a helium bag of 
length 9.65~m to reduce the multiple scattering, the remaining photon beam 
impinged on the analyzer crystal aligned with a goniometer of 20~$\mu$rad 
precision. The number of charged particles coming out of the analyzer crystal 
was counted both by a scintillator (S11) for fast triggering and a solid state 
detector (not shown) for offline analysis. The  S11 scintillator was used to 
detect the  photon conversion into e$^+$e$^-$ pairs at the {\it analyzer} and 
it measured the number of charged particles seen right after the crystal 
{\it analyzer}. For the analysis, we only used events with 2MIPs in S11, as a 
signature for PP events (a minimum ionising particle (MIP) 
leaves a well defined energy in the S11 scintillator, so the selection is for 
e$^+$-e$^-$ events). The photons which did not scatter or interact and the 
electron positron pairs created by the interacting photons continued into a 
magnetic spectrometer.

The pair spectrometer system was introduced next to measure the energy 
spectrum of the photon beam in a multi-photon environment. The dipole analyzer 
magnet (Trim6) of the spectrometer was capable of a maximum beam rigidity of 
0.53~Tm. The tracking elements upstream of the magnet consisted of one drift 
chamber (dch1) for the Ge analyzer and two drift chambers (dch05 and dch1) for 
the diamond analyzer. There were two drift chambers (dch2,dch3) downstream of 
the magnet. The drift chambers measured the horizontal and vertical positions 
of the passing charged particles with a precision of 100\,$\mu$m yielding a 
spectrometer resolution of ${\sigma_p}/{p^2}=0.0012$ with $p$ in units of 
GeV/c.

The calorimeter system measured $E_\gamma^{tot}$, the total radiated energy. 
This was done in a 12-segment array lead-glass calorimeter of 24.6 radiation 
lengths which had a resolution of ${\sigma}/{E}=11.5\%/\sqrt{E}$, with $E$ in 
units of GeV. The central segment of this lead-glass array was used to map 
and align the crystals with an electron beam~\cite{expver}.
A more detailed description of the experimental apparatus is reported 
elsewhere~\cite{mstheses1,mstheses2}.

Various plastic scintillators (Sn or Tn) were used to calibrate the tracking chambers and 
to define different physics triggers. The normalisation event trigger ($norm$) 
consisted of the signal logic combination 
$S1{\cdot}S2{\cdot}\overline{\textrm S3}$ to ensure that an electron is 
incident on the radiator crystal. The scintillator Sc VT rejects radiation 
events coming from the conversion of the tagged photon beam upstream of the 
crystal {\it analyzer}. This trigger is also the minimum bias trigger as it
does not favour any energy, angle or interaction process, but is expected to
enter the experimental set-up correctly.  
The radiation event trigger ($rad$) could then be 
defined as the signal logic combination 
$norm{\cdot}(T1.or.T2){\cdot}\overline{VT}$ indicating that the incoming 
electron has radiated and has been successfully taken out of the photon 
section of the beam line. The pair event trigger ($pair$) was constructed 
as the signal logic combination $rad{\cdot}S11$ to select the events for 
which at least one $e^{+}e^{-}$ pair was created inside the analyzer crystal.

The three different experimental measurements described in this paper 
customised the experimental set-up as follows :
\begin{enumerate}
\item The measurement of the CPP cross section by linearly polarized 
photons on aligned crystals (birefringent effects in CPP) was performed using 
only the radiator and analyzer crystals. Both germanium and diamond analyzers 
were investigated. This measurement also established the new aligned crystal 
polarimetry technique.
\item The investigation of the conversion of linear polarization to circular 
polarization for the CB photons necessitated the inclusion of an additional 
crystal, denoted as the {\it quarter-wave plate}. The crystal polarimetry 
technique was extended to quantify circular polarization as well. The magnet 
B7 served as a sweeping magnet of the particles produced by electromagnetic 
showers in the {\it quarter wave plate}. A solid state detector SSD 
(500\,$\mu$m thick Si crystal, 5$\times$5\,cm$^2$) was placed right after 
the {\it quarter wave crystal} during dedicated runs in order to study the 
shower development.
\item The measurement of the polarization of SOS radiation was done without 
the quarter-wave plate in the system, and in this case the radiator was 
configured to generate SOS radiation.
\end{enumerate}

\begin{table*}[htbp]
\caption{\label{T:crystal}Description of the angle settings for the quarter wave
plate set-up, where $\theta_0$ is the angle between the photon momentum and
crystal axis, $\psi$ is the angle between the photon momentum and the
indicated crystal plane and $\phi$ is the azimuthal angle of the quarter wave
plate relative to the radiator.}
\begin{center}
\begin{tabular}{|c|c|c|c|c|c|}
\hline
Crystal Type   &Purpose     &Axes and Planes      &Orientation            &Thickness     \\
\hline
\hline
Si            &Radiator    &$<$001$>$, (110)      &$\theta_0$=5\,mrad, $\psi_{(110)}$=70\,$\mu$rad               &1.5\,cm      \\
\hline
Si            &Quarter Wave Plate  &$<$110$>$, (110)   &$\theta_0$=2.3\,mrad, $\psi_{(110)}$=0               &10\,cm    \\
\hline
Ge           &Analyzer    &$<$110$>$, (110)   &$\theta_0$=3\,mrad, $\psi_{(110)}$=0               &1\,mm    \\
             & $\eta_1$ measurement&    &   $\phi = \pi/4$, $3\pi/4$               &    \\
             & $\eta_3$ measurement&    &   $\phi = 0$, $\pi/2$               &    \\
\hline
\end{tabular}
\end{center}
\end{table*}

\begin{figure}
\includegraphics[scale=.41]{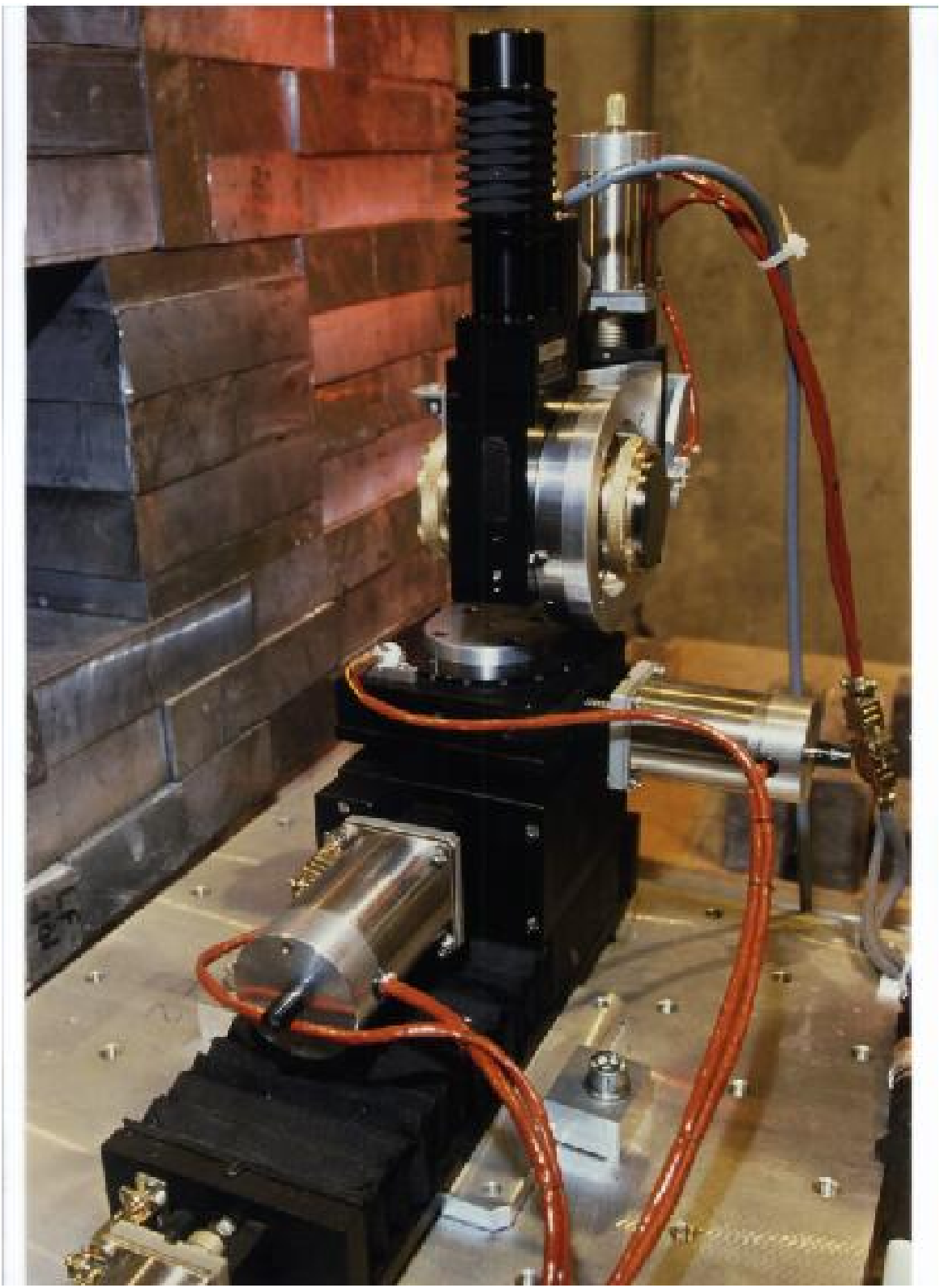}
\caption{\label{F:quarter_photo} {\it Birefringent (quarter wave plate)}
         Si crystal and goniometer.}
\end{figure}

The case of the deployment of the quarter wave plate needs additional 
explanation. A third goniometer controlled the 10\,cm thick Si $<$110$>$ 
crystal, that served as the quarter wave plate. It was located after the 
He-bag. A photograph of the quarter wave plate and the goniometer is shown 
in Fig.~\ref{F:quarter_photo}. The orientation of this crystal relative to 
the photon beam was already discussed in section \ref{subsec:CircPol} 
(see Table~\ref{T:crystal} for a summary of the crystal parameters).

\begin{figure}[htbp]
\includegraphics[width=0.35\textwidth]{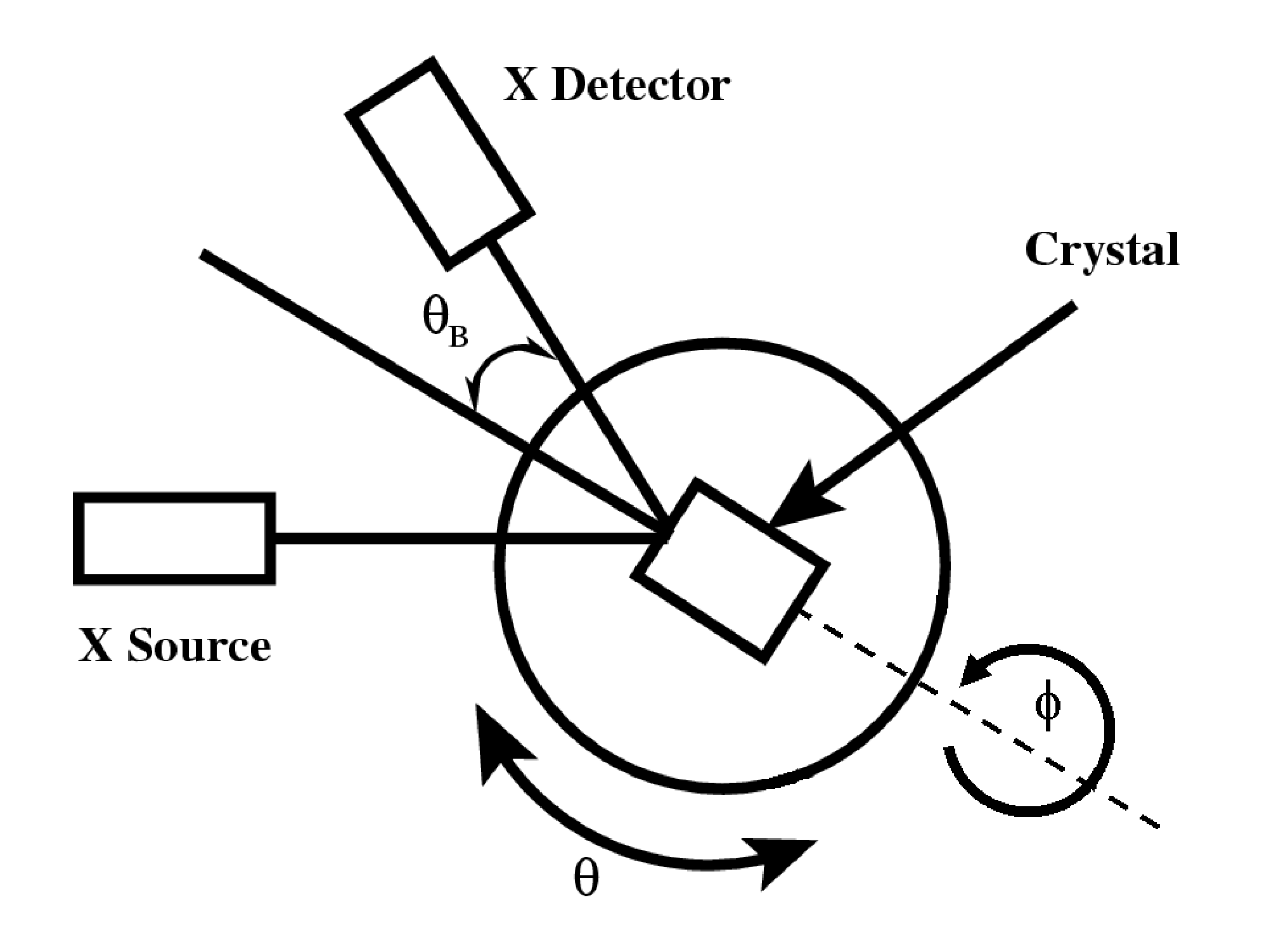}
\includegraphics[width=0.48\textwidth]{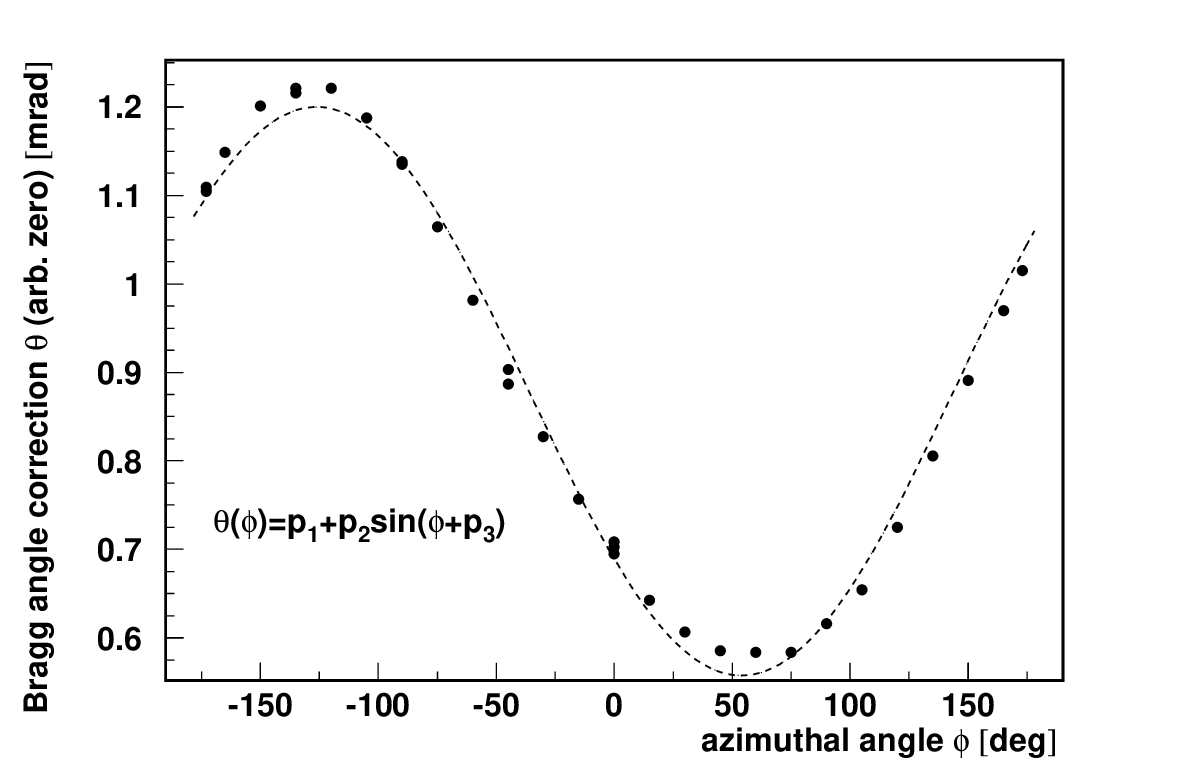}
\caption{\label{F:quarter_alig} Schematic setup (top) and measurements of the 
Bragg peak adjustment angle $\theta$ as a function of the azimuthal rotation 
angle $\phi$ (bottom) for the annular stage alignment 
of the quarter wave plate crystal with X-rays.}
\end{figure}

The axis of the Si crystal was carefully pre-aligned with respect to the axis 
of the azimuthal annular stage that was subsequently mounted into the main 
goniometer.  This pre-alignment procedure was carried out at ESRF, Grenoble. 
A schematic of the alignment setup and the results are shown in 
Fig.~\ref{F:quarter_alig}. An X-ray reflection satisfying the Bragg condition 
was used to monitor the orientation of the $(110)$ crystallographic plane 
which is perpendicular to the $<$110$>$ axis. The crystal was rolled in steps 
using the azimuthal goniometer stage ($\phi$ angle rotation). The $<$110$>$ 
crystallographic axis was slightly misaligned with the crystal physical 
longitudinal axis and therefore also initially slightly misaligned with the 
azimuthal annular stage longitudinal axis. At each azimuthal step the Bragg 
condition had therefore to be recovered by adjustments to the angle of crystal 
face using a second goniometer ($\theta$ angle rotation). The Bragg condition 
was recognised by locating the two points at half-maximum of the Bragg peak. 
From a plot  of the adjustment angle $\theta$ for each step in the roll angle 
$\phi$ of the azimuthal goniometer, the precise offset angles between the 
azimuthal goniometer longitudinal axis and the $<$110$>$ crystallographic 
axis could be obtained. As the thick Si crystal was mounted in the azimuthal 
stage by adjustment screws, the $<$110$>$ crystallographic axis could then be 
brought into coincidence with the longitudinal axis of the azimuthal goniometer.

The data acquisition system consisted of personal computers running  the 
Linux operating system and using in-house developed software to access the 
VME and CAMAC readout crates containing the digitisation modules. The chamber 
signals were read out by VME TDC modules (Caen v767) with 1~ns resolution. 
The scintillator and calorimeter signals were read out with CAMAC ADC modules 
(LRS 2249) with 0.3 pC resolution. The raw data was then stored on DLT tapes 
for offline analysis. The Trigger and DAQ Systems of the experiment are 
described in more detail elsewhere~\cite{Trigger-DAQ}. 

\subsection{\label{subsec:analys}Analysis}

The first step in the offline analysis  was the beam quality cuts, which 
ensured the consistency of various trigger ratios and the initial beam 
position and angles during data taking. Next, to facilitate comparison of the 
experimental results with theoretical predictions, the angular divergence of 
the electron beam was restricted to $\pm $3$\sigma$ from its mean. 
Determination of the electron trajectory and its impact point on the radiator 
were essential for fiducial volume requirements. The radiated photons were 
taken to follow the direction of the initial electron. This is accurate to 
$1/\gamma \approx 5\mu$rad for 100 GeV electrons. To reconstruct the single 
photon energy in each event, only events where a single electron positron 
pair was manifest in the spectrometer volume with the pair energy being the 
same as the photon energy were selected. This subset of pair events were 
further classified into families according to the number of hits on the drift 
chambers of the spectrometer. In our nomenclature, ``122 type'' events are 
clearly the cleanest ones with one hit in the first upstream chamber, and two 
in both the second and third downstream drift chambers. The resulting pair 
production vertex was required to be in the fiducial volume of the analyzer 
crystal. For the case of the diamond analyzer, the additional drift chamber on 
the upstream side ensured a better vertex reconstruction. This in turn 
allowed us to veto the inter-tile events as well as the ones coming from the 
misaligned tile. Quality assessment  of the pair search program was performed 
by a GEANT based Monte Carlo program. This program simulated the effects 
of the detector geometry to understand the precision and efficiency of the 
reconstruction algorithm for each event family. 
Further details of the analysis are recorded in reference~\cite{Analysis}.
 
\begin{table}[htbp]
\caption{\label{tab:data-sets} Different material and angular settings for
      the analyzer crystal used to measure the linear polarization
      components. The angle $\phi$ is defined in the caption to Table I.}
\renewcommand{\extracolsep}{}
\begin{tabular}{|c|c|c|}
\hline
\textbf{analyzer orientation }&
\textbf{\ \ analyzer\ \ }&
\textbf{measured polariza-}\\
\textbf{($\phi$)}&
\textbf{type}&
\textbf{tion component}\\
\hline
\hline
$0,\frac{\pi }{2},\pi ,\frac{3\pi }{2}$& Ge& $\eta _{3}$\\ \hline
$\frac{\pi }{4},\frac{3\pi }{4},\frac{5\pi }{4},\frac{7\pi }{4}$& Ge&
$\eta _{1}$\\ \hline
$0,\frac{\pi }{2}$& Diamond& $\eta _{3}$\\ \hline
\end{tabular}
\end{table}

During the data taking, to obtain the parallel and perpendicular 
configurations, the angular settings of the radiator crystal (hence the 
direction of linear polarization of the photon beam) were kept constant. Only 
the analyzer crystal was rotated in a rolling motion around its symmetry axis.
Therefore to measure the magnitude of the $\eta _{3}$ ($\eta _{1}$) component 
of the polarization,  analyzer orientations separated by $\pi /2$ starting 
from 0 ($\pi /4$) were compared. To reduce the systematic errors, (especially 
in the case of the Ge crystal where the analyzing power is smaller), all 
relevant angles on the analyzer crystal were utilised for polarization 
measurements, as presented in Table~\ref{tab:data-sets}. Other sources of 
systematic errors were the uncertainty in the crystal angles, the photon 
tagging and the pair reconstruction efficiencies obtained from Monte Carlo studies.

\subsection{\label{subsec:CB-valid}CB Validation}

\begin{figure}[htbp]
\includegraphics[width=0.45\textwidth]{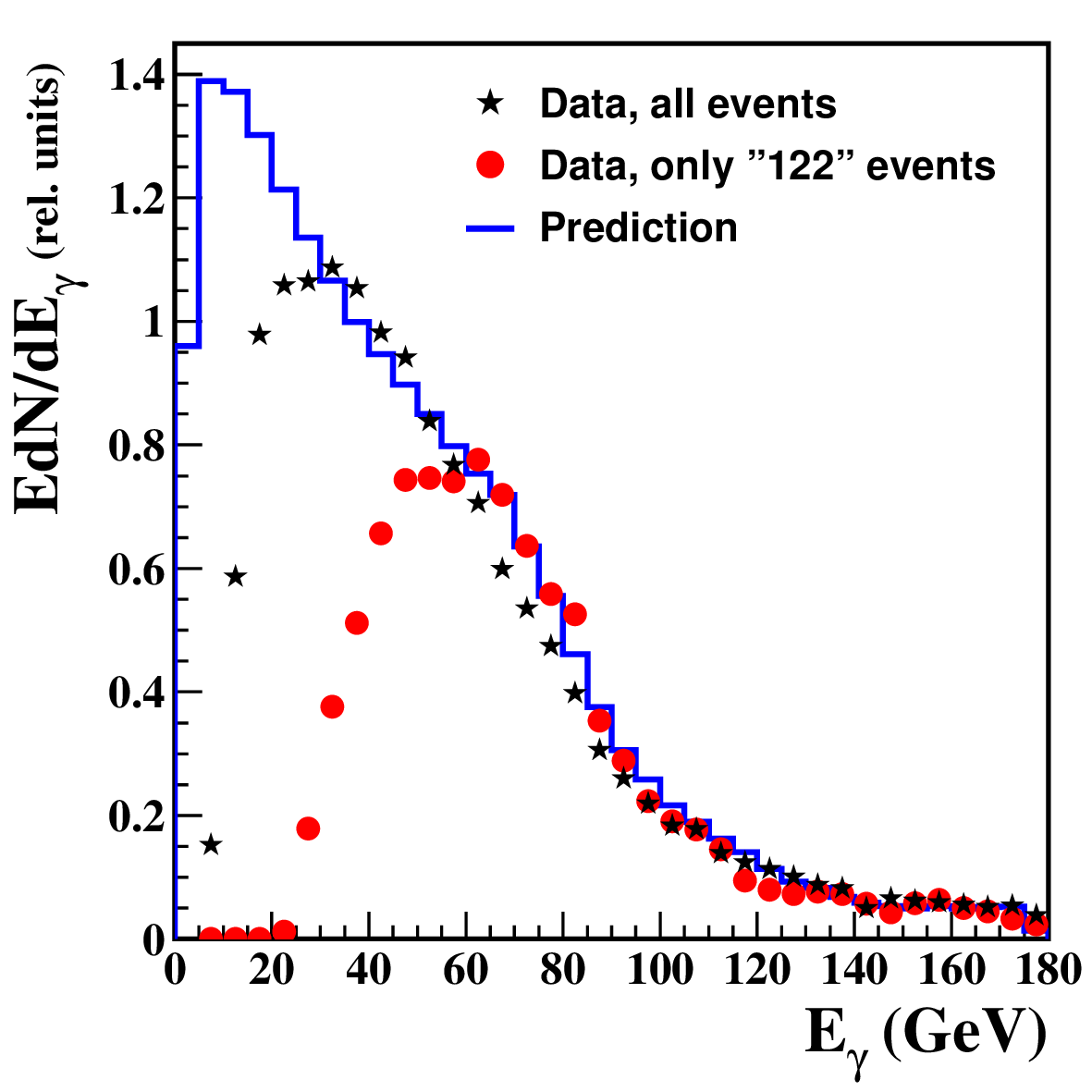}
\caption{\label{fig:single-photons} Monte Carlo predictions for the single photon
   spectrum,  compared with data using all events (stars) and only `122
   type' events (circles).}
\end{figure}

The angular settings of the radiator crystal were verified by inference from 
the data. The single photon intensity spectrum presented in 
Fig.~\ref{fig:single-photons} contains two different event selections 
superimposed on the Monte Carlo prediction. The geometrical acceptance of the 
spectrometer for "122 type" events, named after their reconstruction in drift chambers,
has a relatively high threshold of 30 GeV, as seen from 
the Fig.~\ref{fig:single-photons}. The large number of low energy photons is due to the thickness of the target (1.5cm) and the small angle of incidence (70urad).

\begin{figure}[htbp]
\includegraphics[width=0.48\textwidth]{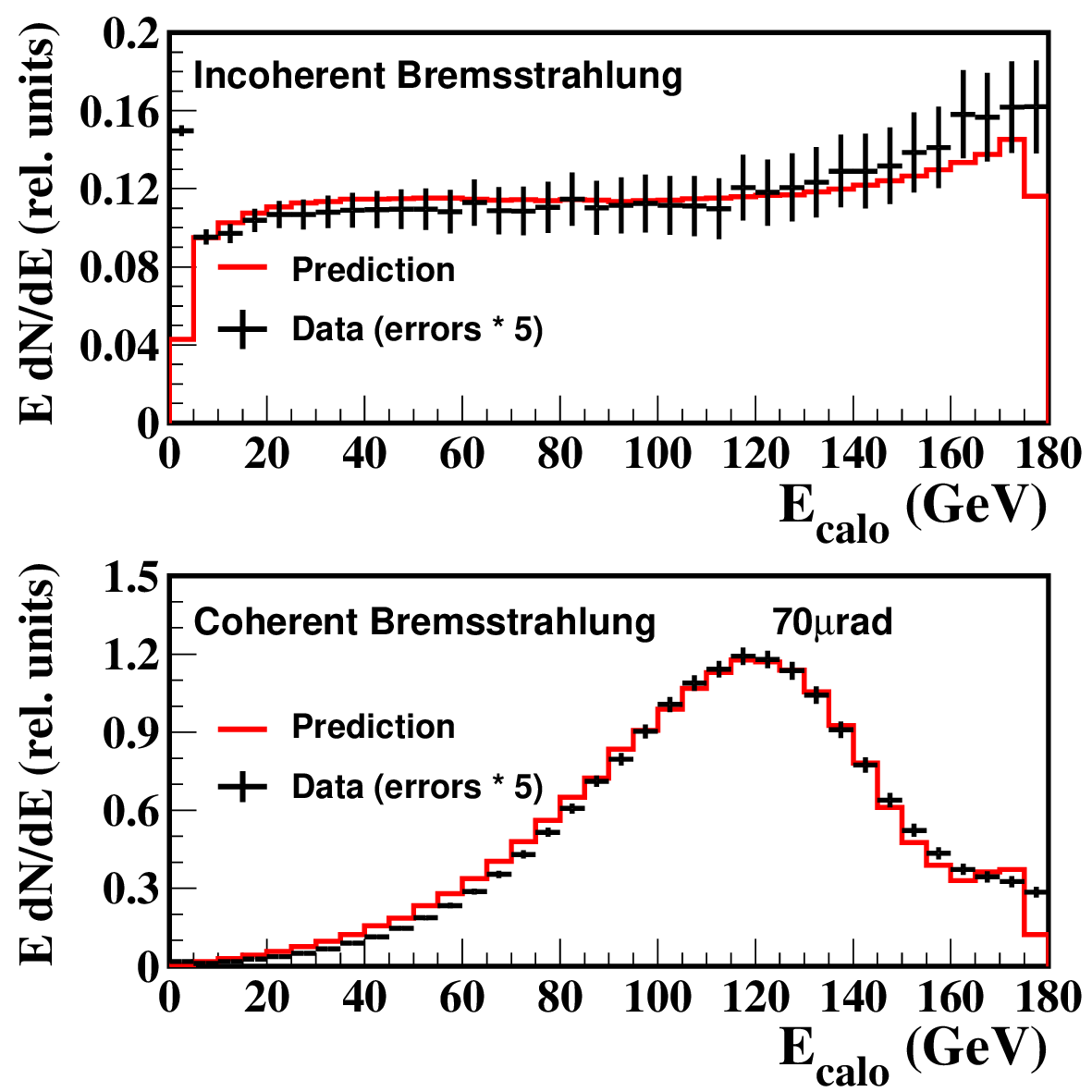}
\caption{\label{fig:pileup} 
Power spectrum for incoherent (top) and for coherent bremsstrahlung 
(bottom) as a function of the total energy $E_{calo}$ 
detected in the calorimeter.
Vertical axis: Power (rel. units). The statistical errors on the data
are enhanced by a factor of five to increase visibility.}
\end{figure}

An independent method of verifying the CB settings is looking at the total 
electromagnetic radiation from the radiator crystal. 
Fig. ~\ref{fig:pileup} shows the total power spectrum (E dN/dE) as a function of the
energy measured in the calorimeter for the radiator crystal not-aligned
(Fig. ~\ref{fig:pileup}(top)) and aligned (Fig.~\ref{fig:pileup}(bottom)).
In terms of the 
radiation intensity spectrum, an unaligned crystal is identical to an 
amorphous material. This radiation is called ICB and it can be approximated by 
the familiar Bethe-Heitler formula \cite{BH}. The increase in the CB radiation 
intensity spectrum is usually reported with respect to the IB spectrum. This 
ratio, called the ``enhancement'', is presented in Fig.~\ref{fig:enhancement} 
together with Monte Carlo prediction for CB angle at 70\,$\mu$rad. The agreement of 
the data with the enhancement prediction is  remarkable. The offline analysis 
could therefore be used to monitor the angular settings of the radiator in time 
steps, to ensure the crystal angular settings did not drift during the 
measurement.

\section{\label{sec:res}Results}

\subsection{\label{subsec:CB-polarim}Birefringence in CPP and a new crystal 
polarimetry}

\begin{figure}[htbp]
\includegraphics[width=0.45\textwidth]{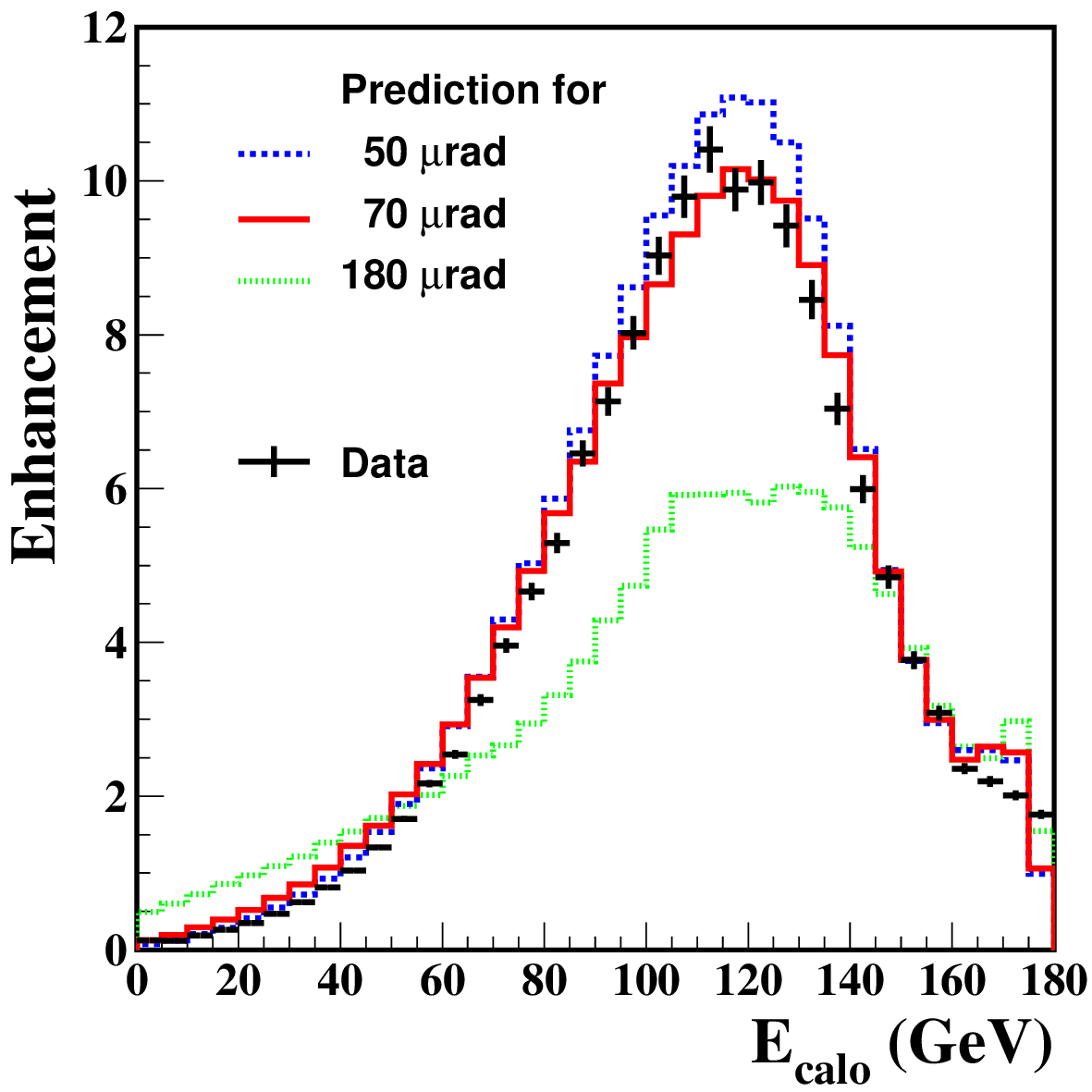}
\caption{\label{fig:enhancement} Enhancement of CB radiation data
compared to  Monte Carlo predictions. Note the sensitivity of the cross section to
small changes in the angular setting of the crystal.}
\end{figure}

\begin{figure}[htbp]
\includegraphics[scale=0.46]{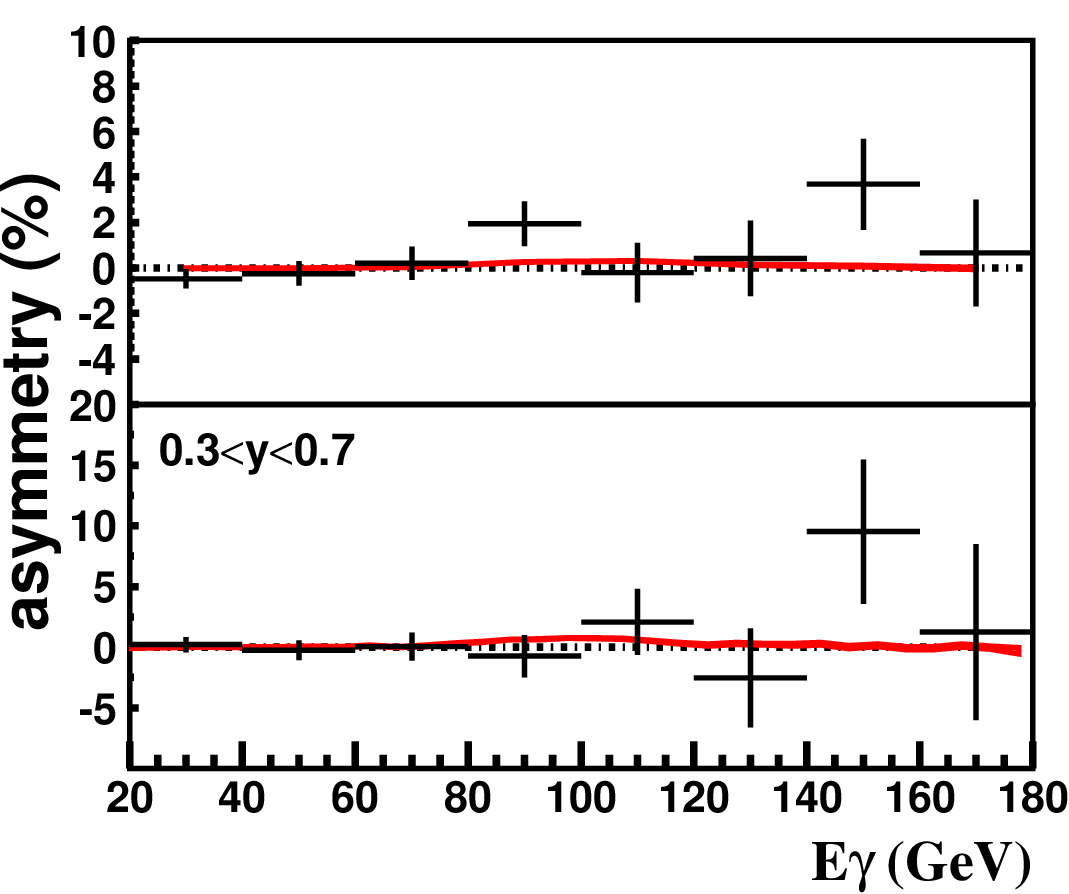}
\caption{\label{fig:eta1-rslt-ge} Asymmetry to determine $\eta_1$
component of the photon polarization with the Ge analyzer. The data at
roll angles $\pi/4 + 5\pi/4$ are compared to $3\pi/4 + 7\pi/4$ {\it
without (top)} and {\it with (bottom)} the quasi-symmetrical pair
selection.}
\end{figure}

The orientation of the radiator crystal could be accurately determined by 
comparison of the predicted and measured CB enhancement 
(Fig.~\ref{fig:enhancement}). The predicted and measured asymmetries for both 
linear polarization components: $\eta_1$ and $\eta_3$ could then be 
confidently compared. Using all events, as well as events passing the 
quasi-symmetrical pairs selection criteria, we see that, as expected, the 
asymmetry in Fig.~\ref{fig:eta1-rslt-ge} is consistent with zero yielding a 
vanishingly small $\eta_1$ component of the polarization.

\begin{figure}[htbp]
\includegraphics[scale=0.46]{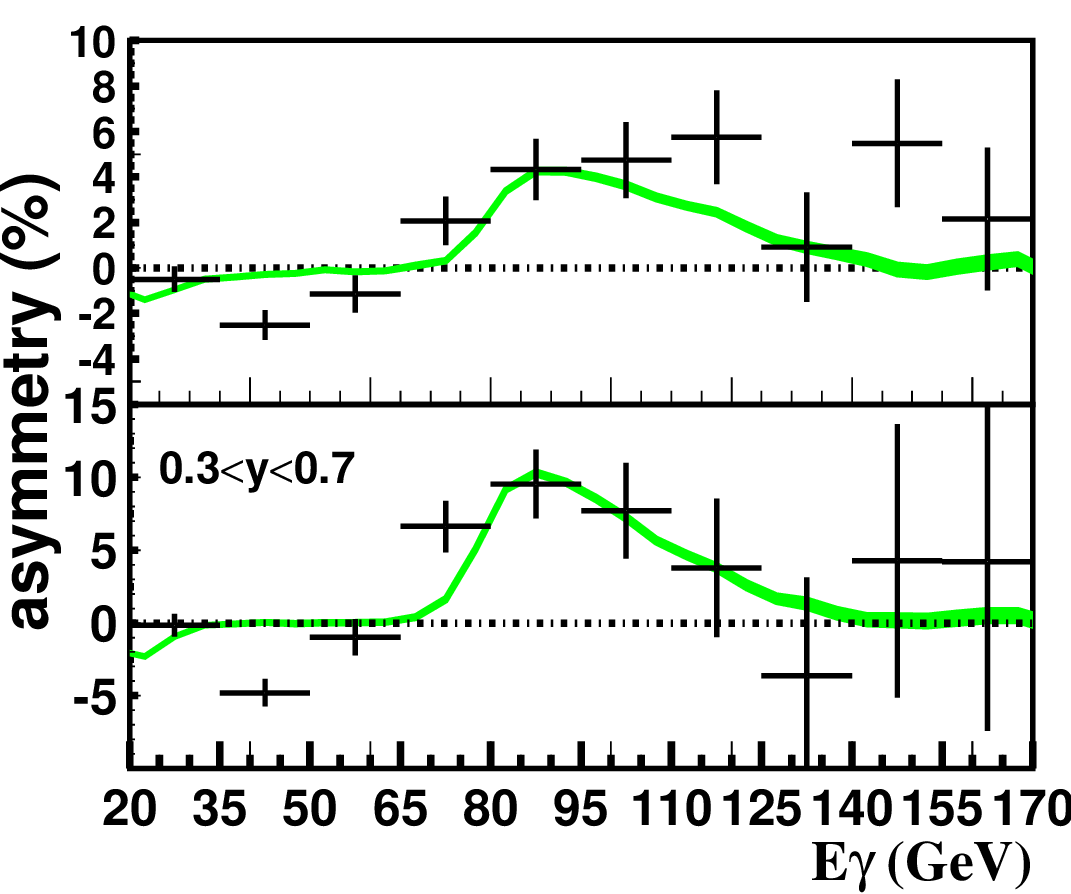}
\caption{\label{fig:eta3-rslt-ge} Asymmetry to determine the $\eta_3$
component of the CB photon polarization with the Ge analyzer. Measurements
{\it without (TOP) } and {\it with (BOTTOM) } the quasi-symmetrical pair
selection at roll angles $0 + \pi$ are compared to those at roll angles
$\pi/2 + 3\pi/2$.}
\end{figure}

\begin{figure}[htbp]
\includegraphics[scale=0.46]{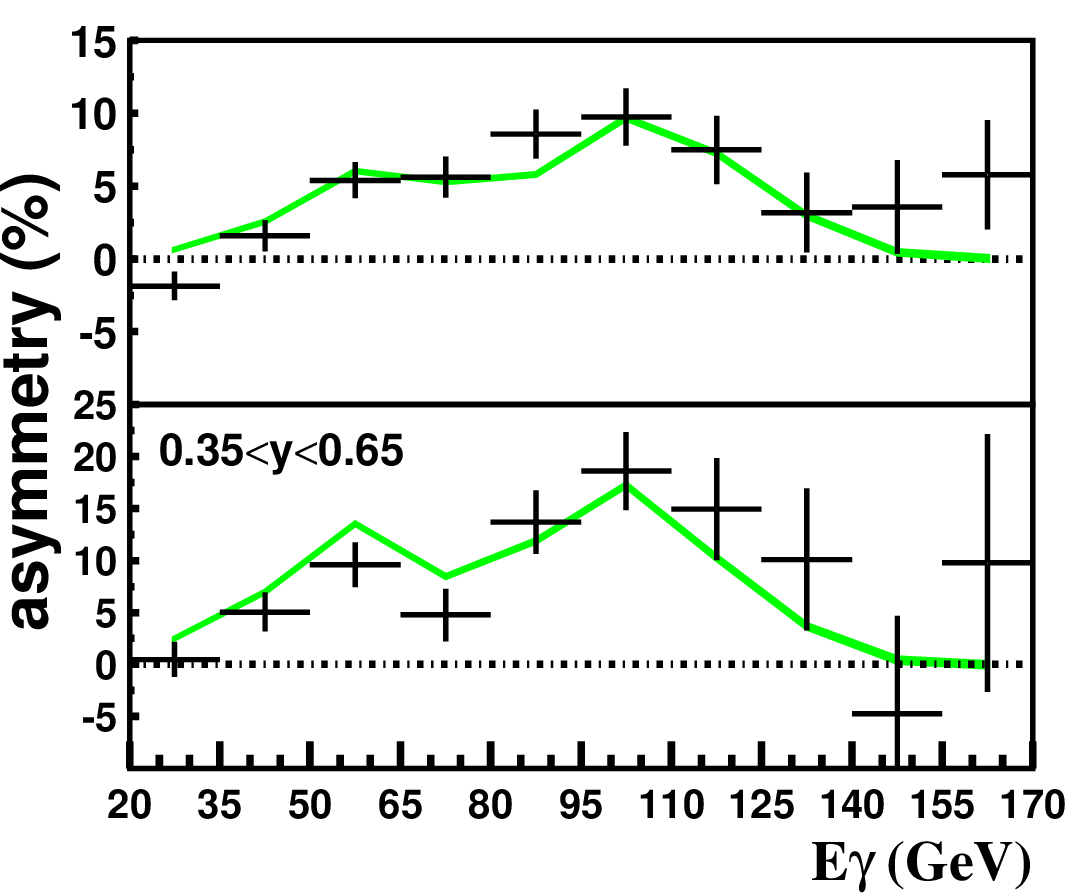}
\caption{\label{fig:eta3-rslt-di} Asymmetry measurements {\it without
(TOP)} and {\it with (BOTTOM)} the quasi-symmetrical pair selection to
determine $\eta_3$ component of the photon polarization with the {\it
diamond} analyzer (Cf. Table~\ref{tab:data-sets}).}
\end{figure}

The measured asymmetry in the induced polarization direction ($\eta_3$) is 
presented in Fig.~\ref{fig:eta3-rslt-ge} without and with the $y$-cut using 
the Ge analyzer crystal. The solid line represents the Monte Carlo predictions without 
any smearing effects considered in the spectrometer. The lower plot  represents the increase in 
the asymmetry due to quasi-symmetrical pairs together with the statistical 
error associated with this increase. It thus confirms the non statistical 
source of the asymmetry increase in the 70-110~GeV range. The same 
polarization as measured by the diamond analyzer is given in 
Fig.~\ref{fig:eta3-rslt-di}. The top and middle plots show again the asymmetry 
measurements as compared to the Monte Carlo predictions without any smearing, and the 
lower plot shows the increase in the asymmetry due to the $y$-cut. Comparing 
figures~\ref{fig:eta3-rslt-ge} and \ref{fig:eta3-rslt-di}, we conclude that 
the multi-tile synthetic diamond crystal is a better choice than the Ge 
crystal as an analyzer, since for the same photon polarization the former 
yields a larger asymmetry and thus enables a more precise measurement. The 
diamond analyzer also allowed the measurement of the  photon polarization in 
the 30-70~GeV range, since it has some, albeit small, analyzing power at these 
energies.

The theoretical predictions are based both on the calculation of the energy 
dependent polarization of photons produced by coherent bremsstrahlung and the 
polarization dependence of coherent pair production, also as a function of 
incident energy. Thus the polarization sensitive versions of both CB and 
CPP are needed together in the theoretical calculation that predicts the 
measured asymmetry. The theoretical calculation combines the coherent and 
quasiclassical theories of radiation and pair production, in a Monte Carlo 
approach that can describe real beams with finite divergence. The agreement of 
this combined theory with the measured data is remarkable. It is clear that 
for the energy range of 30-170 GeV and the incident angle phase space of this 
study that the theory is sufficiently reliable and well understood to support 
the development of applications of crystals as polarimetry devices. The 
calculation of the resolving power (R in equation~\ref{eq:asy-def}) is 
therefore reliable for the energy and angle regimes discussed in the 
introduction. The asymmetry measurements therefore correspond to a measurement 
of the induced polarization for CB for  $\eta_3$ shown in 
Fig.~\ref{fig:pol-predict}. This has a maximum of 57\% at 70\,GeV.

\subsection{\label{sec:exp_quarter}Conversion of linear to circular 
polarization}

In this part of the experiment, the  {\it quarter wave plate} crystal was 
introduced between the {\it radiator} and {\it analyzer} crystals. The linear 
polarization measurements mentioned in the previous section were extended to 
allow the measurement of circular polarization. This measurement is related to 
a reduction in linear polarization and the conservation of polarization. The 
theoretical background was described in section~\ref{subsec:CircPol}. It is 
noted that the expected and measured single photon spectrum for the chosen CB 
parameters for the {\it radiator} as shown in Fig.~\ref{fig:single-photons} 
are in good agreement. The  expected polarization for the same set of 
parameters has already been given in Fig.~\ref{fig:pol-predict} as a function 
of the single photon energy. As shown in the previous section, this 
polarization has been confirmed for the $\eta_1$ and $\eta_3$ components. This 
indicates that the linearly polarized CB photon beam is well understood.

\begin{figure}[t]
\includegraphics[scale=0.558]{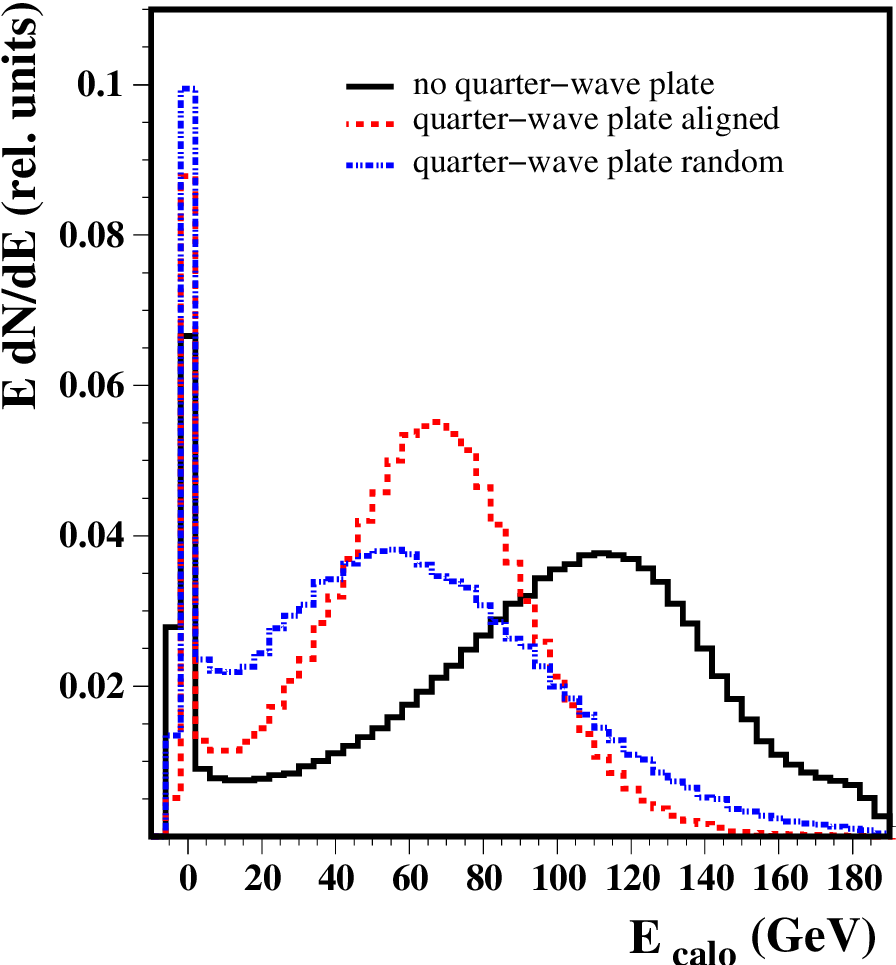}
\caption{\label{F:l4-calo} Power spectrum of the 
photon beam measured in the calorimeter.
Without the quarter wave plate crystal the predicted CB spectrum is
observed (solid line). With the aligned quarte wave plate inserted
the spectrum shifts to lower energies (dashed line). And with the
quarter wave plate in random orientation we see more spread in the
shifted spectrum.}
\end{figure}

Any change in the single photon spectrum after adding the {\it quarter wave 
crystal} will reflect how the incoming photons are absorbed or transformed by 
it. As $\eta_1$ was found to be consistent with zero before the {\it quarter 
wave crystal}, any nonzero value observed after it is a reflection of 
birefringent effects of the crystal.

Detailed theoretical calculations and simulations have been done to choose the 
crystal type, orientation and optimal thickness for the {\it quarter wave 
crystal}, leading to the choice of a 10\,cm thick Si crystal as discussed 
above. The analysis took into account the real experimental parameters 
including the angular spread of the incident photon beam, the generation of 
secondary particles, multiple Coulomb scattering, and  all particles produced 
by electromagnetic showers were also taken into account. In the simulation we 
assume the angular spread of the photons with energies between 70-100\,GeV to 
be about $\sim$60\,$\mu$rad and $\sim$45\,$\mu$rad in horizontal and vertical 
planes, respectively, as measured from the data. The calculations also include 
the polarization transformation part for the surviving photon beam, resulting 
in elliptical polarization.

Fig.~\ref{F:l4-calo} shows the photon beam power spectrum measured with the LG 
electromagnetic calorimeter. The calorimeter sees all the surviving photons 
radiated by the parent electron. By comparing the spectrum with the 
{\it quarter wave crystal} at random and/or aligned with the case in which 
there is no {\it quarter wave crystal}, we can see that the {\it quarter wave 
plate} consumes a significant amount of the beam. This  causes the peak energy
of the pileup spectrum to be reduced by at least 50\,GeV. However, it is also 
clear that the energy of the photons absorbed by the {\it quarter wave crystal} 
depends on its alignment condition.

\begin{figure}[ht]
\includegraphics[scale=0.433]{na59-joint-fig19}
\caption{\label{F:absor} Absorption probability found from the ratio of
     the single photon spectrum for the data with and without the {\it
     quarter wave crystal}.}
\end{figure}

As already mentioned in section III.C, the prediction is that 
only 17-20 \% of the photons is expected to survive
in the energy region of interest.  This is confirmed by the data, see 
Fig.~\ref{F:absor}. In addition, it is clear that the survival probability is 
also energy dependent as expected.

Another consequence of adding the {\it quarter wave crystal} is a significant 
increase in the photon multiplicity of an event. For example, we expect an 
average multiplicity of three photons per electron for the nominal 
{\it radiator} settings. By analyzing the correlation between the calorimeter 
spectrum and the single photon spectrum, we can conclude that the majority of 
these photons have energies $<$5~GeV, and that the calorimeter events at high 
energies are dominated by a single high energy photon and not due to the 
pileup of many low energy photons. As a consequence, the measurement of the 
Stokes parameters in the high energy range can be performed by measuring the 
asymmetry using either the calorimeter or the pair spectrometer.

\begin{figure}[htbp]
\includegraphics[width=0.48\textwidth]{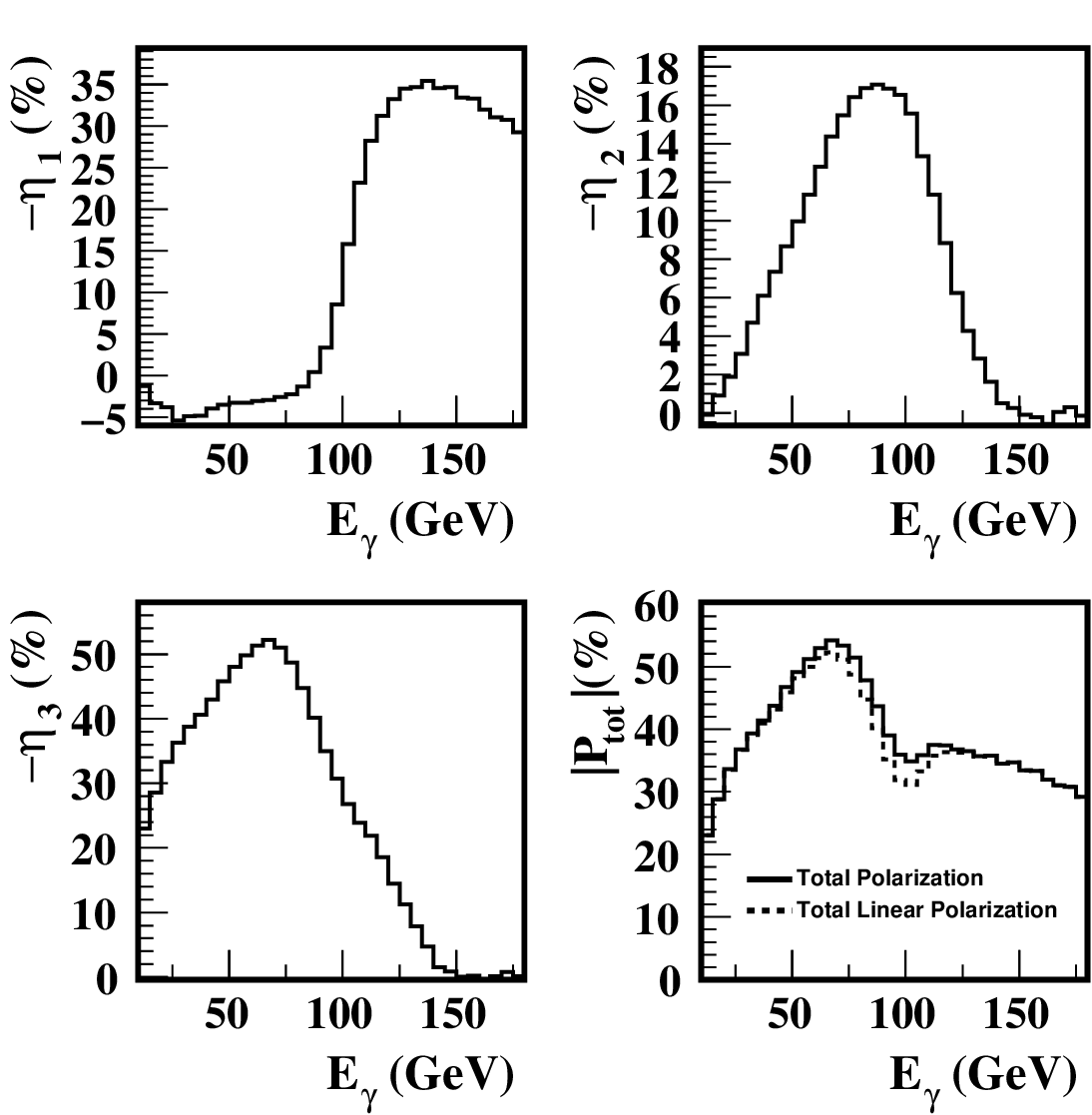}
\caption{\label{F:l4-stokes} Stokes parameters after the {\it quarter wave
    crystal}, assuming as input the values given in
    Fig.~\ref{fig:pol-predict}.}
\end{figure}

The expected Stokes parameters and the total polarization of the photons after 
the {\it quarter wave crystal} are given in Fig.~\ref{F:l4-stokes}. As shown, 
the expected value of the $\eta_3$ Stokes parameter decreases from 
36$\%$ to 30$\%$ around 100\,GeV. This difference should be seen in the PP 
asymmetry. The expected degree of circular polarization is of the order of 
$\sim$16$\%$ at the same energy. In Fig.~\ref{F:l4-stokes}, we expect an 
interesting increase of up to a factor of seven for the $\eta_1$ Stokes 
parameter in the same energy region.  This phenomenon was also predicted by 
Cabibbo~\cite{cabibbo2}, the unpolarized photon beam traversing the aligned 
crystal becomes linearly polarized. This follows from the fact that the 
high-energy photons are mainly affected by the PP process. This cross section 
depends on the polarization direction of the photons with respect to the 
plane passing through the crystal axis and the photon momentum (polarization 
plane). Thus, the photon beam penetrating the oriented single crystal feels 
the anisotropy of the medium. For the experimental verification of this 
phenomenon with photon beams at energies of 9.5\,GeV and 16\,GeV, 
see~\cite{berger, eisele}. In the high energy region $>$100\,GeV the 
difference between the PP cross sections parallel and perpendicular to the 
polarization plane is large. Since the photon beam can be regarded as a 
combination of two independent beams polarized parallel and perpendicular 
with respect to the reaction plane, one of the components will be absorbed 
to a greater degree than the other one, and the remaining beam becomes 
partially linearly polarized.

\begin{figure}[htbp]
\includegraphics[scale=0.433]{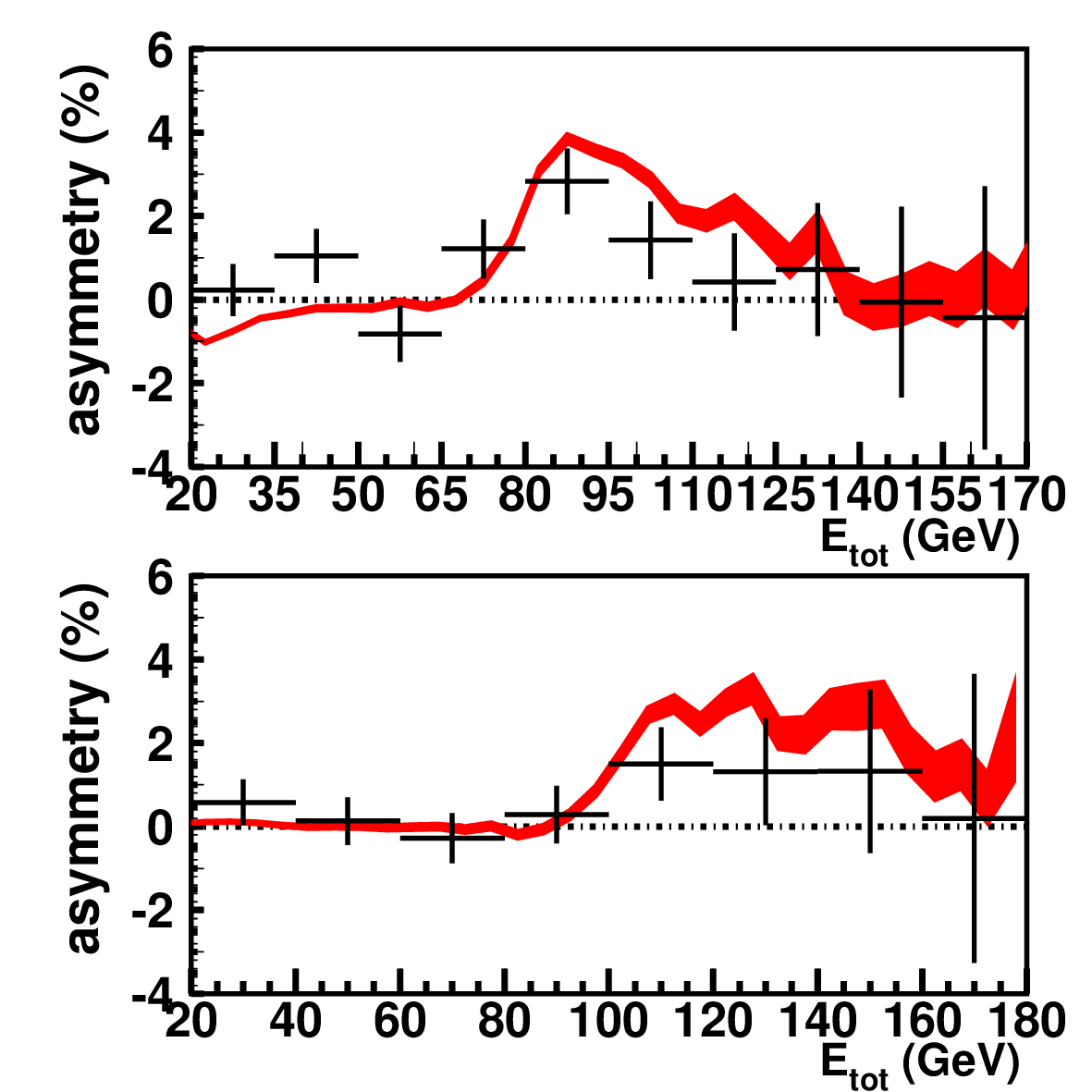}
\caption{ \label{F:l4-asy1} Asymmetry measured with the calorimeter. The
    results reflects the changes in $\eta_3$\,(top) and generation of
    $\eta_1$\,(bottom) due to the presence of the {\it quarter wave
    crystal}.}
\end{figure}

\begin{figure}[ht]
\includegraphics[scale=0.46]{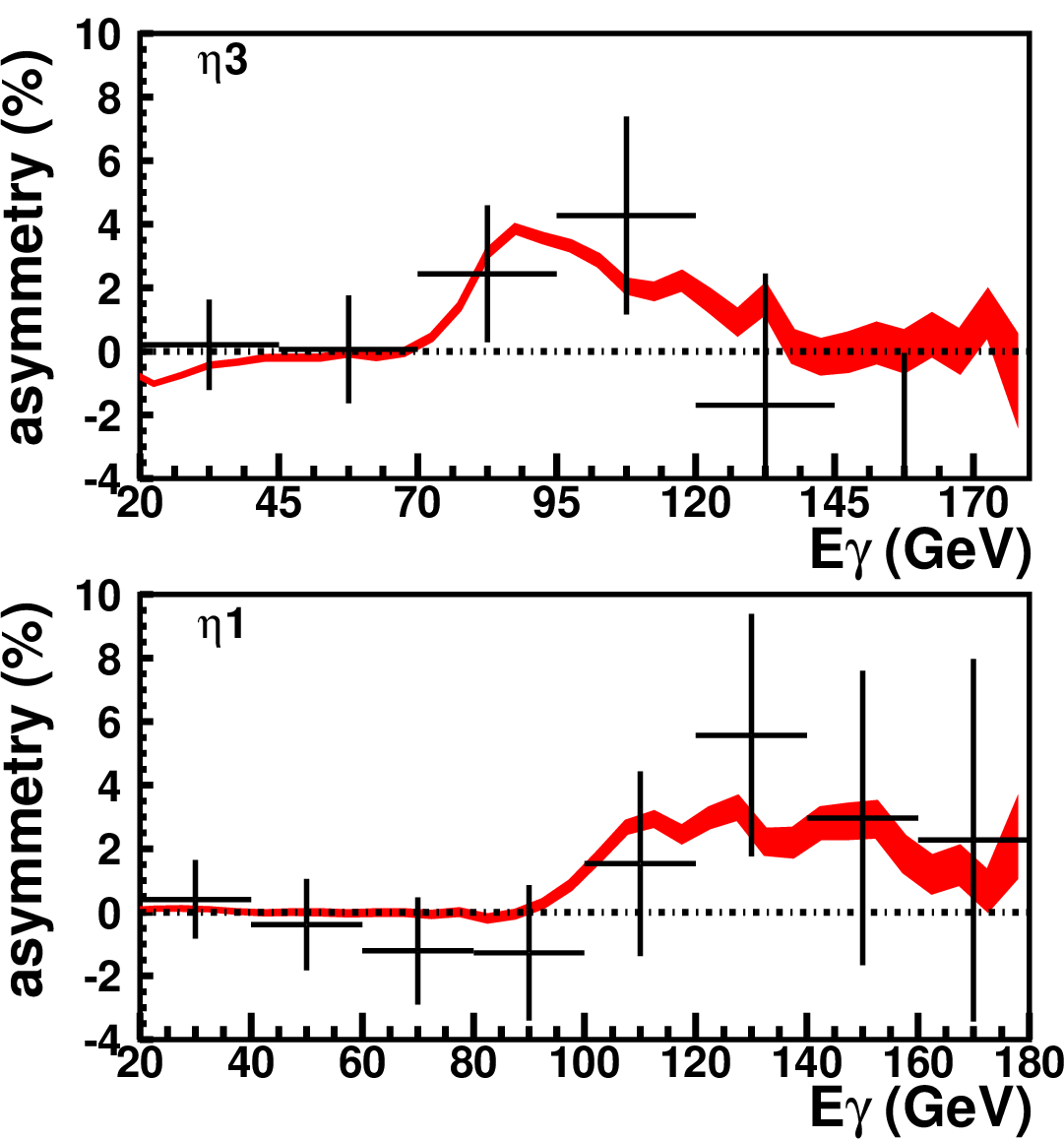}
\caption{\label{F:l4-asy2} Asymmetry measured with the 
    Ge analyzer (top) and the pair spectrometer (bottom).
    The data sample  with a fully reconstructed single e$^+$e$^-$ pair is
    ten times smaller than the total data sample with at least one pair
    passing all the data quality cuts and with a 2MIPs cut (see section IV-A) in S11.}
\end{figure}

The measured asymmetries using the 
calorimeter are given in Fig.~\ref{F:l4-asy1} and again using the 
pair-spectrometer in Fig.~\ref{F:l4-asy2}. In order to reduce systematic 
uncertainties, the angular settings of the {\it radiator} crystal (hence the 
direction of linear polarization of photon beam) were kept constant, and only 
the {\it analyzer} crystal was rolled around its symmetry axis to obtain the 
parallel and perpendicular configurations. Therefore, to measure the 
polarization of the $\eta_{3}$ ($\eta_{1}$) component, the asymmetry between 
the 0 ($\pi /4$) and $\pi /2$~($3\pi /4$) {\it analyzer} orientations were 
used.

As shown in these figures, the measured asymmetries are in agreement with the 
predicted polarization for the chosen Ge {\it analyzer} crystal 
setting~\cite{overview}. For the Stokes parameter $\eta_{3}$, the measured 
asymmetry after the {\it quarter wave crystal} is about 2.9$\pm$0.7\% in the 
energy range  between 80-100\,GeV. The estimated analyzing power $R$ for the 
Ge~{\it analyzer} in the same energy range is about 10\%~\cite{overview}. 
Using the equation~(\ref{eq:asy-def}) one can estimate the measured Stokes 
parameter $\eta_3$ after the {\it quarter wave crystal}. Thus, the measured 
Stokes parameter is $\eta_3$=28$\pm$7\% (see Fig.~\ref{F:l4-stokes}). For the 
Stokes parameter $\eta_3$, the measured asymmetry without the {\it quarter wave 
crystal} in the same energy range was found to be 4.7$\pm$1.7\%, 
(see~\cite{overview}). This corresponds to a measured value of 
$\eta_3$=44$\pm$11\%, which is also consistent with the theoretically expected 
value of $\eta_3$, see Fig.~\ref{fig:pol-predict}.

Similar calculations may be done for the Stokes parameter $\eta_1$. If we make 
a weighted average for the asymmetry values between 20 and 100 GeV, where we 
expect no asymmetry, we obtain a value of 0.19$\pm$0.3\%. Above 100\,GeV we 
expect a small asymmetry, where we measured (1.4$\pm$0.7)\% between 100 and 
180\,GeV.
                                                                                
Using the equation~(\ref{eq:no7}) one can now find the measured circular 
polarization degree which is equal $\eta_2$=21$\pm$11\%. This is consistent 
with the predicted value of 16\%.

The statistical significance of the result was estimated using the F-test to 
evaluate the confidence level associated with distinguishing between two 
different statistical distributions. The first distribution was formed by the 
variance of the energy dependent data for the experimental circular 
polarization with respect to the theoretical prediction displayed in 
Fig.~\ref{F:l4-stokes}. The second distribution was formed by the variance of 
the same data to the null hypothesis prediction of no circular polarization. 
Limiting the test to the region where the crystal polarimeter has analyzing 
power, and also to the region where the circular polarimetry technique of 
equation \ref{eq:no7} has efficiency (80 - 110GeV), then we find a confidence 
limit of 73\% for the observation of circular polarization.

\subsection{\label{subsec:sos-pol}Polarization measurement of SOS radiation}

This third section of the experiment can be divided in two parts: (1) 
production of the photon beam by the photon radiation from the 178 GeV 
electron beam in the Si radiator oriented in the SOS mode and (2) measurement 
of the linear polarization using diamond crystals as analyzers. Prior to the 
experiment Monte Carlo simulations were used to estimate the photon yield, the radiated 
energy, and the linear polarization of the photon beam and we optimized the 
orientation of the crystal radiator. The Monte Carlo calculations also included the 
crystal analyzer to estimate the asymmetry of the e$^+$e$^-$ pair production. 
The simulations further included the angular divergence of the electron beam, 
the spread of 1\% in the beam energy, and the generation of the 
electromagnetic shower that develops in oriented crystals. To optimize the 
processing time of the Monte Carlo simulation, energy cuts of 5 GeV for electrons and 
of 500 MeV for photons were applied.

\subsubsection{\label{subsubsec:photon-beam}Photon Beam}
The SOS photons were produced as discussed in section~\ref{subsec:SoS-prod} 
and Fig.~\ref{F:Strak-1b} displays a theoretical calculations of the various 
components of the photon power yield per unit of thickness of radiator crystal 
as used in the experiment. However there are several consequences for the 
photon spectrum due to the use of a 1.5~cm thick crystal. For the chosen 
orientation of the relatively thick Si crystal, the emission of mainly low 
energy photons from planar coherent bremsstrahlung results in a total average 
photon multiplicity above 15. The most probable radiative energy loss of the 
178~GeV electrons is expected to be 80\%. The beam energy decreases 
significantly as the electrons traverse the crystal. The peak energy of both 
SOS and PC radiation also decreases with the decrease in electron energy. 
Consequently, the SOS radiation spectrum is not peaked at the energy for a 
thin radiator, but becomes a smooth energy distribution.

\begin{figure}[hbpt]
\includegraphics[scale=0.433]{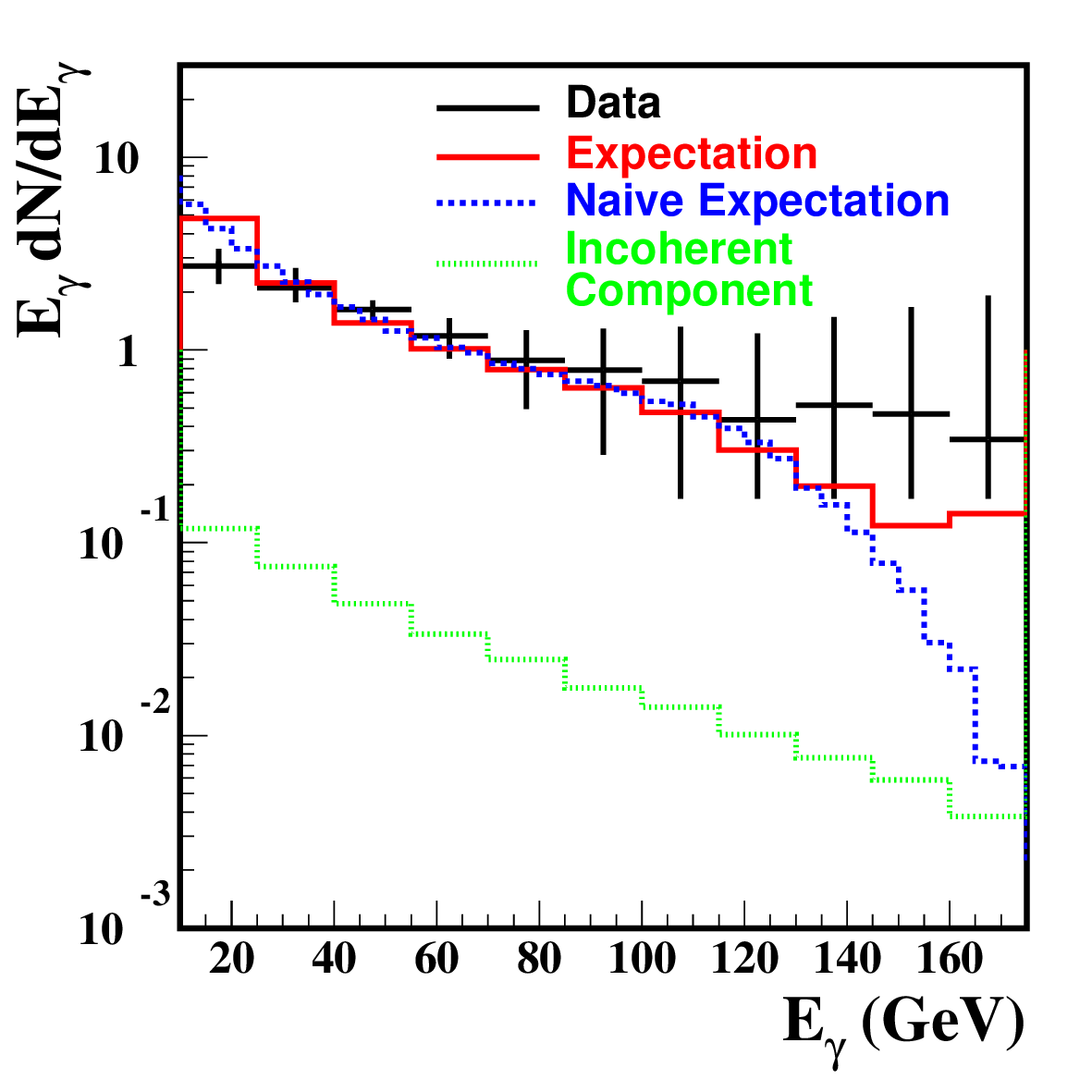}
\caption{\label{F:sps} Photon power yield, $E_\gamma dN/dE_\gamma$, as a
    function of the energy $E_\gamma$ of individual photons radiated by an
    electron beam of 178\,GeV in the SOS-aligned 1.5\,cm Si crystal. The
    black crosses are the measurements with the pair spectrometer, the
    vertical lines represent the errors including the uncertainty in the
    acceptance of the spectrometer. The (red solid) histogram represent
    the Monte Carlo prediction for our experimental conditions. The (green dotted)
    represent the small contribution due to incoherent interactions. For
    completeness, we also show the theoretical predictions if the
    experimental effects are ignored  (blue dashed).}
\end{figure}

Clearly, many 
electrons may pass through the crystal without emitting SOS radiation and 
still lose a large fraction of their energy due to PC and ICB. Hard photons 
emitted in the first part of the crystal that convert in the later part do 
not contribute anymore to the high energy part of the photon spectrum. A full 
Monte Carlo calculation is necessary to propagate the predicted photon yield with a 
thin crystal, as shown in Fig.~\ref{F:Strak-1b} for 178 GeV electrons, to the 
current case with a 1.5~cm thick crystal. This has been implemented for the 
measured photon spectrum shown in Fig.~\ref{F:sps}. We see that the 
measured SOS photon spectrum shows a smoothly decreasing distribution. 
Consequently, the high energy radiation is emitted essentially in the very 
first part of the crystal, while soft photons will be emitted along the full 
length of the crystal. This effect has been observed 
previously~\cite{kirsebom01}. The low energy region of the photon spectrum is 
especially saturated, due to the abundant production of low energy photons. 
Above 25 GeV however, there is satisfactory agreement with the theoretical 
Monte Carlo prediction, which includes the effects mentioned above.

\begin{figure}[htbp]
\includegraphics[scale=0.42]{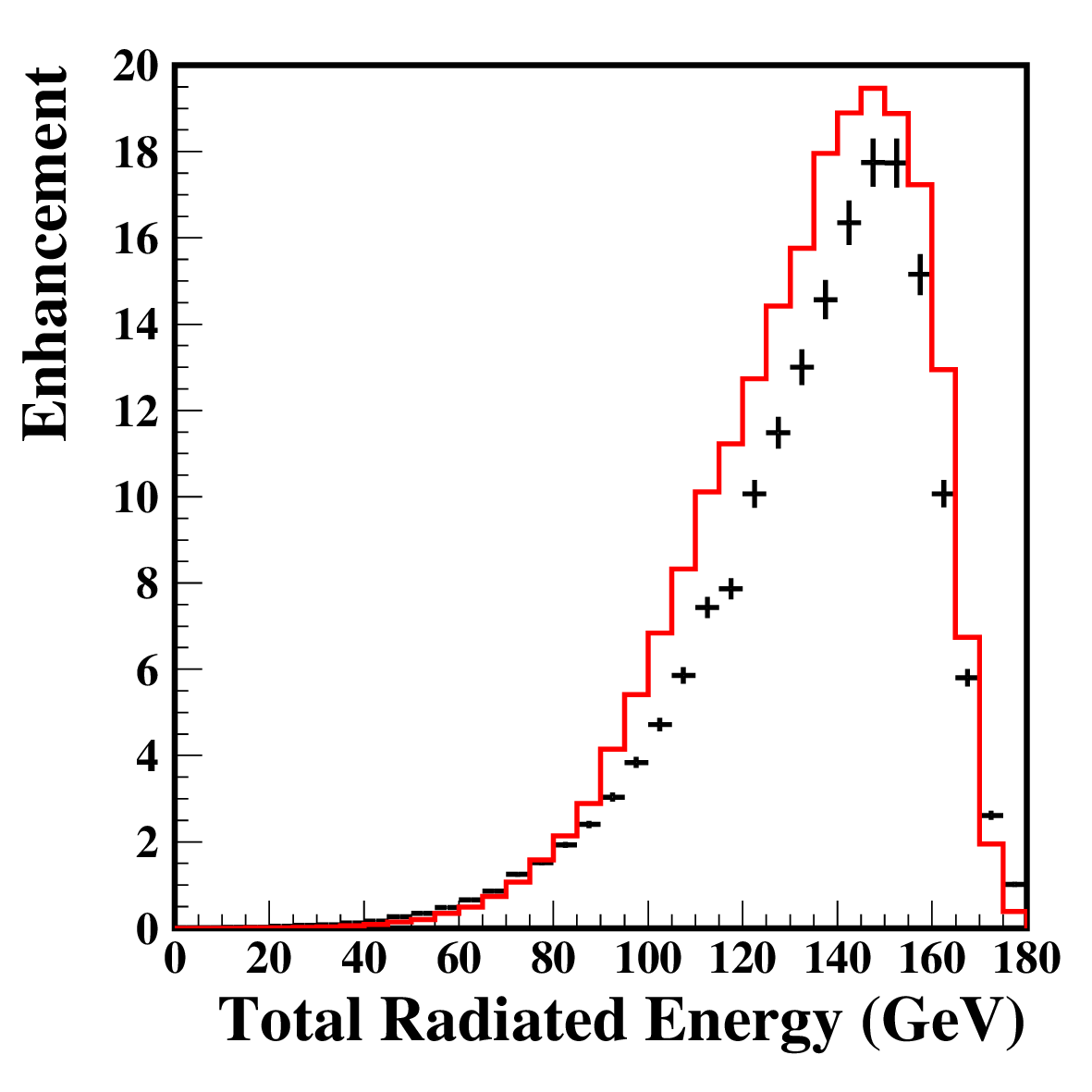}
\caption{\label{F:enh} Enhancement of the intensity with respect to the
     Bethe-Heitler\,(ICB) prediction for randomly oriented polycrystalline
     Si as a function of the total radiated energy $E_{tot}$ in the
     SOS-aligned Si crystal by 178\,GeV electrons. The black crosses are
     the measurements and the red histogram represent the Monte Carlo prediction.}
\end{figure}

The enhancement of the emission probability compared to the ICB prediction is 
given in Fig.~\ref{F:enh} as a function of the total radiated energy as 
measured in the calorimeter. The maximal enhancement is about a factor of 18 
at 150 GeV and corresponds well with the predicted maximum of about 20 at 
148~GeV. This is a multi-photon spectrum measured with the photon calorimeter. 
The peak of radiated energy is situated at 150 GeV, which means that each 
electron lost about 80\% of its initial energy due to the large thickness of 
the radiator. This means that the effective radiation length of the oriented 
single crystal is several times shorter in comparison with the amorphous 
target. The low energy region is depleted due to the pile-up of several photons.

\indent

\begin{figure}[htbp]
\includegraphics[scale=0.433]{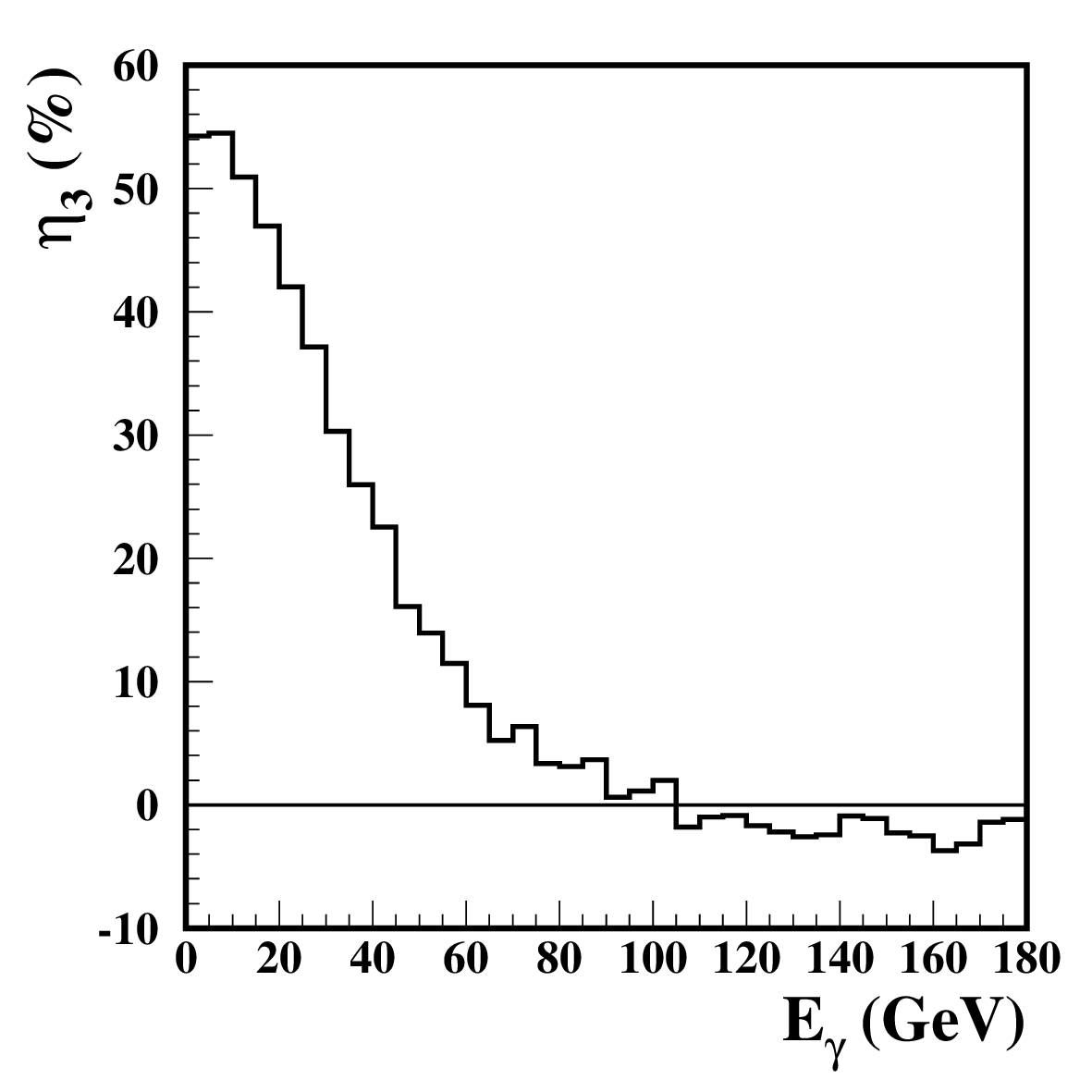}
\caption{\label{F:SOS-pol} Expected linear polarization as a function of
    the energy $E_\gamma$ of the photons produced in the SOS-aligned Si
    crystal by 178\,GeV electrons.}
\end{figure}

The radiator angular settings were chosen to have the total linear polarization 
from the SOS radiation  purely along $\eta _{3}$, that is $\eta _{1}=0$. The 
$\eta _{2}$ component is also zero because the electron beam is unpolarized. 
The expected $\eta _{3}$ (linear polarization) component of the  polarization 
shown is in Fig.~\ref{F:SOS-pol} as a function of photon energy. It is well 
known that channelling radiation in single crystals is linearly 
polarized~\cite{adishchev,vorobyov} and the low energy photons up to 70\,GeV 
are also predicted to be linearly polarized in the Monte Carlo simulations. High energy 
photons are predicted to have a small degree of polarization.

\subsubsection{\label{subsubsec:asym-measure}Asymmetry Measurement}

\begin{figure}[ht]
\includegraphics[scale=0.433]{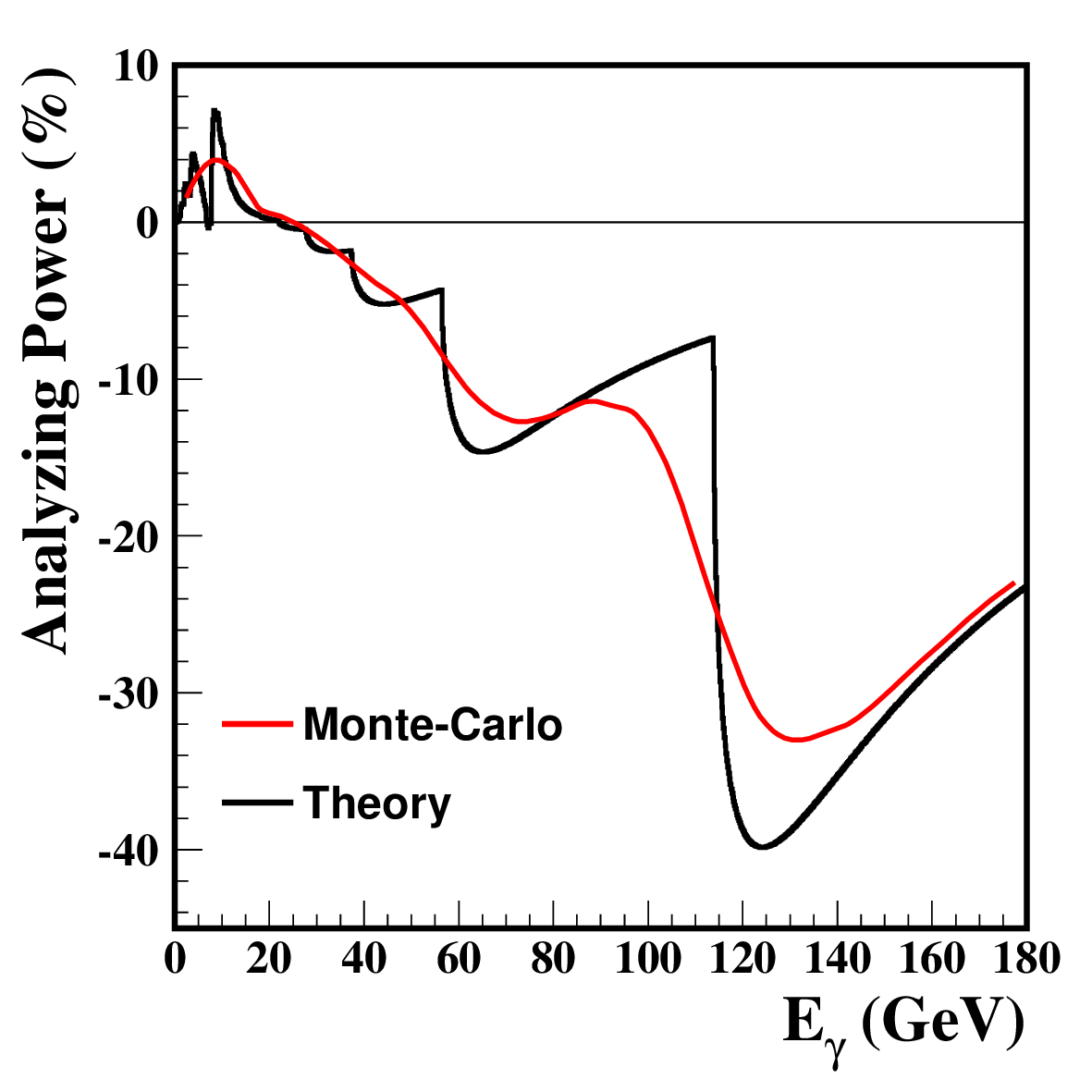}
\caption{\label{F:anpow} Analyzing power $R$ with the aligned diamond
    crystal as a function of the photon energy $E_\gamma$ (black curve)
    for an ideal photon beam without angular divergence and (red curve)
    for the Monte Carlo simulation of photons with the beam conditions in
    the actual experiment.}
\end{figure}

The polarization measurement was made as explained in 
section~\ref{subsec:Xtal-pol}. A multi-tile synthetic diamond crystal was used 
as an analyzer oriented with the photon beam at 6.2~mrad to the 
$\langle 100 \rangle$ axis and at 465~$\mu$rad from the $(110)$ plane. This 
configuration is predicted to have a maximal analyzing power for a photon 
energy of 125 GeV as is shown in Fig.~\ref{F:anpow}. The predicted analyzing 
power in the high energy peak region is about 30\%.

\begin{figure}[htbp]
\includegraphics[width=0.48\textwidth]{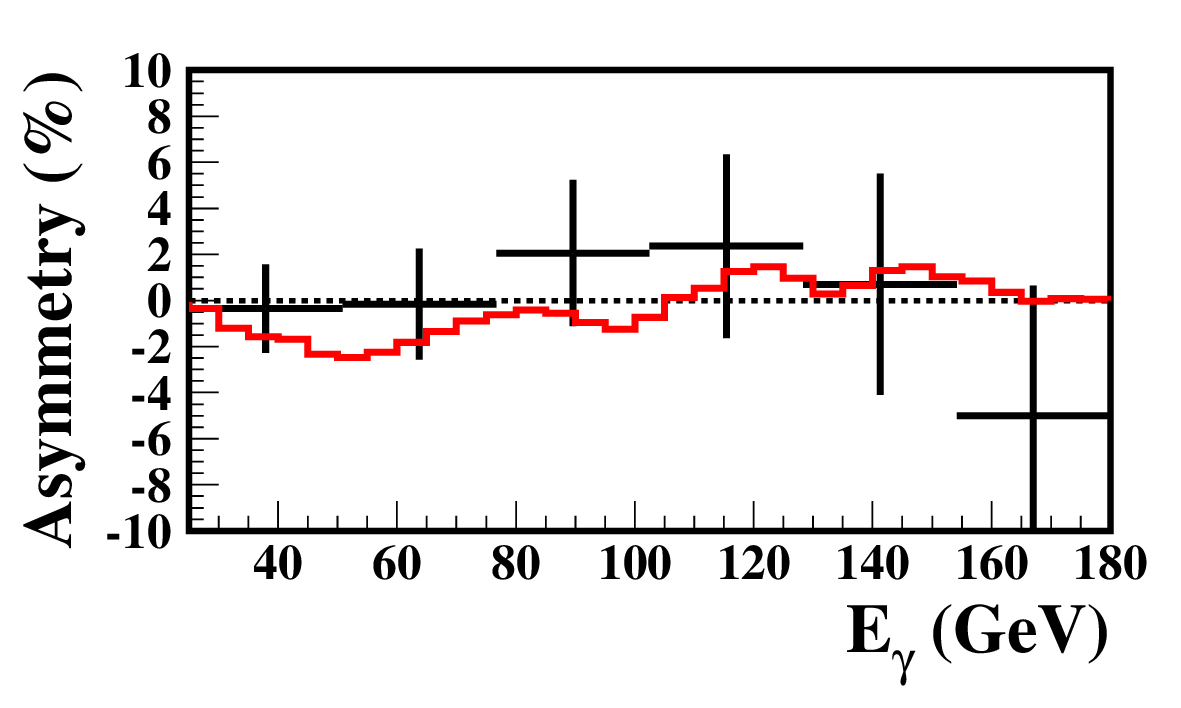}
\caption{\label{F:asy} Asymmetry of the e$^+$e$^-$ pair production in the
     aligned diamond crystal as a function of the photon energy $E_\gamma$
     which is measured to determine the $P_1$ component of the photon
     polarization in the SOS-aligned Si crystal by 178\,GeV electrons. The
     black crosses are the measurements and the red histogram represent
     the Monte Carlo prediction.}
\end{figure}

The measured asymmetry and the predicted asymmetry are shown in 
Fig.~\ref{F:asy}. One can see that the measured asymmetry is consistent with 
zero over the whole photon energy range. For the photon energy range of 
100-155~GeV we find less than 5\% asymmetry at 90\% confidence level using the 
F-test of significance. The null result is expected to be reliable as the 
correct operation of the polarimeter has been confirmed in the same beam-time 
in measurements of the polarization of CB radiation as described in 
section~\ref{subsec:CB-polarim}. Note, that the expected asymmetry is small, 
especially in the high energy range of 120-140 GeV, where the analyzing power 
is large, see Fig.~(\ref{F:anpow}). This corresponds to the expected small 
linear polarization in the high energy range, see Fig.~(\ref{F:SOS-pol}).

In contrast to the result of a previous experiment \cite{kirsebom99}, our 
results are consistent with calculations that predict a polarization of only a 
few percent in the high energy photon peak for the SOS orientation. The 
analyzing power of the diamond analyzer crystal in the previous 
experiment's~\cite{kirsebom99} setup peaked in the photon energy range of 
20-40~GeV where a high degree of linear polarization is expected. However for that 
experiment in the high energy photon region we expect a small analyzing power 
of about 2-3\%, also following recent calculation~\cite{simon,strakh}. The 
constant asymmetry measured in a previous experiment~\cite{kirsebom99} over 
the whole range of total radiated energy may therefore not be due to the 
contribution of the high energy photons. 

From Fig.~\ref{F:SOS-pol} one can expect a large linear polarization for 
photons in the low energy range of 20-50~GeV. However, the analyzing power 
was optimized for an photon energy of 125~GeV and is small in the region where 
we expect a large polarization. A different choice of orientation of the 
analyzer crystal can move the analyzing power peak to the low energy range and 
may be used to measure the linear polarization in the low energy range.

\section{Conclusions}

\subsection{Birefringence in CPP and a new crystal polarimetry}

Our results presented in the previous section show the feasibility of aligned crystals 
for linearly polarized high energy photon beams. From the experimental point 
of view, for the creation of a photon beam with a predictable spectrum the 
crucial components are (i) high precision goniometers to align the radiator 
crystal with respect to the electron beam and (ii) tracking chambers to monitor 
the incident angles of the electron beam on the crystal surface. The 
predictability of the photon energy and polarization is a good asset for 
designing future beamlines and experiments. The results also establish the 
applicability of aligned crystals as polarimeters for an accurate measurement 
of the photon polarization at high energies. The important aspects are the 
selection of the analyzer material and the utilization of quasi-symmetric 
pairs. The use of available synthetic diamond as analyzer crystal is found to 
be very promising due to its durability and high analyzing power.

The pair spectrometer enables us to do asymmetry measurements for single 
photons in a multi-photon environment. If the photon multiplicity is low, as 
it would be for laser generated beams with $E > 10$\,GeV, then a simple 
multiplicity detector can be used to replace the more complex pair 
spectrometer. This is especially the case for a multiplicity detector which is 
energy selective. CB events with high photon multiplicity are known to be 
dominated by a single high energy photon and accompanied by multiple low energy 
photons.

The crystal polarimetry technique developed here will also be applicable in 
high energy photon beamlines as a fast monitoring tool. For example, in a 
future $\gamma \gamma$ or $e \gamma$ collider quasi-online monitoring of the 
photon beam polarization could be achieved using this crystal polarimetry 
method. In the most competitive designs of such colliders~\cite{crapxing}, 
the photon beam after the interaction region is transferred to a beam dump, 
hence the destructive nature of the crystal polarimetry technique does not 
constitute an impediment for its utilization.

\subsection{Conversion of linear to circular polarization}

The experimental results of this section show that coherence effects in single 
crystals can be used to transform linear polarization of high-energy photons 
into circular polarization and vice versa. Thus, it seems possible to produce 
circularly polarized photon beams with energies above 100GeV at secondary 
(unpolarized) electron beams at high energy proton accelerators. The 
birefringent effects becomes more pronounced at higher photon energy, which 
allows for thinner crystals with higher transmittance.

Diamond will be more efficient than silicon as quarter wave plates, and a 2~cm 
thick diamond crystal will have a transmission probability of about 80\% for 
100~GeV photons. A diamond array of 0.4~cm thickness was produced and aligned 
for our experiments, where we used it as a linear polarization analyzer, see 
Fig~\ref{fig:diamonds}.

A robust measurement of the circular polarization would involve measurement of
the decay asymmetry of $\rho$-mesons produced in and behind the birefringent 
Si crystal. An alternative method was used here.  The  
aligned pair production crystal was used as an analyzer, and realistic theoretical 
calculations describe very well (i) the radiated photon spectrum from the 
aligned radiator and (ii) the pair production asymmetries in the aligned 
analyzer with and without the birefringent Si crystal in the photon beam. In 
view of this good agreement all the predicted effects, including the 
birefringent effect, seem to be confirmed by the present measurements. 
Measurements of the charged particle multiplicity with depleted Si detectors 
show a large sensitivity to crystal alignments and can be used to control the 
alignment of crystals and the photon polarization in a future polarimeter 
set-up.

\subsection{Polarization measurement of SOS radiation }

We have performed an investigation of both enhancement and polarization of 
photons emitted in the so called SOS radiation mode. This is a special case 
of coherent bremsstrahlung for multi-hundred GeV electrons incident on 
oriented crystalline targets, where the hardness and the enhancement of the 
photon spectrum is more favourable than in the normal CB case. The experimental
set-up had the capacity to deal with the relatively high photon multiplicity 
and single photon spectra were measured. This is very important in view of the 
fact that there are several production mechanisms for multiple photons
which have different radiation characteristics. We have confirmed the single 
photon nature of the hard photon peak produced in SOS radiation.

The issue of the polarization of the SOS photons had previously not been 
settled conclusively. The results of an earlier experiment~\cite{kirsebom99} 
indicated that a large polarization might be obtained for the high energy SOS 
photons. Our experimental results show that the high energy photons emitted 
by electrons passing through the Si crystal radiator oriented in the SOS mode 
have a linear polarization smaller than 20\% at a confidence level of 90\%.

Since the previous experiments, the theoretical situation for the polarization 
of hard SOS photons has also become clearer. Our results also confirm these 
recent calculations which predict that the linear polarization of high energy 
photons created in SOS orientation of the crystal is small compared to the 
polarization obtained with the PE orientation.

Photon emission by electrons traversing single crystals oriented in the SOS 
orientation has interesting peculiarities since three different radiation 
processes are involved: (1) incoherent bremsstrahlung, (2) channeling 
radiation and (3) coherent bremsstrahlung induced by the periodic structure of 
the atomic strings in the crystal that are crossed by the electron. The recent 
calculations have taken these three processes into account and predict around 
a 5\% polarization for the high energy SOS photons. This prediction is 
consistent with our zero polarization result from the asymmetry measurement 
of single photons with energies above 100 GeV.

\begin{acknowledgments}

We dedicate this work to the memory of Friedel Sellschop. We express our
gratitude to CNRS, Grenoble for the crystal alignment and Messers DeBeers
Corporation for providing the high quality synthetic diamonds.  We are
grateful for the help and support of N. Doble, K. Elsener and H. Wahl. It
is a pleasure to thank the technical staff of the participating
laboratories and universities for their efforts in the construction and
operation of the experiment.

This research was partially supported by the Illinois Consortium for
Accelerator Research, agreement number~228-1001. UIU acknowledges support
from the Danish Natural Science research council. The support of the 
National Research Foundation of South Africa is also acknowledged.

\end{acknowledgments}

\bibliography{na59-joint}

\end{document}